\documentclass{aa}  
\usepackage{graphicx}
\usepackage{txfonts}
\usepackage[switch,columnwise]{lineno}
\usepackage[unicode=true,psdextra]{hyperref}
\begin{document}

\title{Optical spectroscopy of blazars for the Cherenkov Telescope Array - III}

\author{F. D'Ammando          \inst{1}
        \and
        P. Goldoni \inst{2}
        \and
        W. Max-Moerbeck \inst{3}
        \and
        J. Becerra González \inst{4,5}
        \and
        E. Kasai \inst{6}
        \and
        D. A. Williams \inst{7}
        \and
        N. Alvarez-Crespo \inst{8}
        \and
        M. Backes \inst{6,9} 
        \and
        U. Barres de Almeida  \inst{10}
        \and
        C. Boisson \inst{11}
        \and
        G. Cotter \inst{12}
        \and
        V. Fallah Ramazani \inst{13} 
        \and
        O. Hervet \inst{7}
        \and
        E. Lindfors \inst{14}
        \and
        D. Mukhi-Nilo \inst{15}
        \and
        S. Pita \inst{16}
        \and
        M. Splettstoesser \inst{7} 
        \and
        B. van Soelen \inst{17}
        }

\institute{
INAF - Istituto di Radioastronomia, Via Gobetti 101, I-40129 Bologna, Italy 
\and Universit\'{e} Paris Cit\'{e}, CNRS, CEA, Astroparticule et Cosmologie, F-75013 Paris, France
\and Departamento de Astronom\'{i}a, Universidad de Chile, Camino El Observatorio 1515, Las Condes, Santiago, Chile
\and
Universidad de La Laguna (ULL), Departamento de Astrof\'isica, E-38206 La Laguna, Tenerife, Spain
\and
Instituto de Astrof\'isica de Canarias (IAC), E-38200 La Laguna, Tenerife, Spain
\and Department of Physics, Chemistry \& Material Science, University of Namibia, Private Bag 13301, Windhoek, Namibia
\and 
Santa Cruz Institute for Particle Physics and Department of Physics, University of California, Santa Cruz, CA 95064, USA
\and
IPARCOS and Department of EMFTEL, Universidad Complutense de Madrid, E-28040 Madrid, Spain
\and
Centre for Space Research, North-West University, Potchefstroom 2520, South Africa
\and Centro Brasileiro de Pesquisas Fisicas (CBPF), Rua Dr. Xavier Sigaud 150 - Urca, Rio de Janeiro 22290-180, Brazil
\and Laboratoire Univers et Th\'{e}ories, Observatoire de Paris, Universit\'{e} PSL, Universit\'{e} Paris Cit\'{e}, CNRS, F-92190 Meudon, France
\and Oxford Astrophysics, University of Oxford, Denys Wilkinson Building, Keble Road, Oxford, OX1 3RH, United Kingdom
\and Ruhr-Universität Bochum, Fakultät für Physik und Astronomie, Astronomisches Institut (AIRUB), 44801 Bochum, Germany
\and Finnish Centre for Astronomy with ESO, FINCA, University of Turku, Turku, FI-20014 Finland
\and Instituto de Astrof\'{i}sica, Facultad de Fis\'{i}ca, Pontificia Universidad Cat\'{o}lica de Chile, Av. Vicuña Mackenna 4860, Macul, Santiago, Chile
\and Universit\'{e} Paris Cit\'{e}, CNRS, Astroparticule et Cosmologie, F-75013 Paris, France
\and Department of Physics, University of the Free State, Bloemfontein 9300, South Africa}

\date{Received XXX; accepted XXX}


\abstract
   {Blazars, which include BL Lacs and flat-spectrum radio quasars (FSRQs), represent the brightest persistent extragalactic sources in the high-energy (HE; 10 MeV -- 100 GeV) and very-high-energy (VHE; $E$ $>$ 100 GeV) $\gamma$-ray sky. Due to their almost featureless optical/UV spectra, it is challenging to measure the redshifts of BL Lacs. As a result, about 50\% of $\gamma$-ray BL Lacs lack a firm measurement of this property, which is  fundamental for population studies, indirect estimates of the extragalactic background light (EBL), and  fundamental physics probes (e.g. searches for Lorentz-invariance violation or axion-like particles).
   }
   {This paper is the third in a series of papers aimed at determining the redshift of a sample of blazars selected as prime targets for future 
   observations with the next generation, ground-based VHE $\gamma$-ray astronomy observatory, Cherenkov Telescope Array Observatory (CTAO). The accurate determination of the redshift of these objects is an important aid in source selection and planning of future CTAO observations.
   }
   {Promising targets were selected following a sample selection obtained with Monte Carlo simulations of CTAO observations. The selected targets were expected to be detectable with CTAO in observations of 30 hours or less. We performed deep spectroscopic observations of 41 of these blazars using the Keck II,  Lick, SALT, GTC, and ESO/VLT telescopes. We carefully searched  for spectral lines in the spectra and whenever features of the host galaxy were detected, we attempted to model the  properties of the host galaxy. The magnitudes of the targets at the time of the observations were also compared to their long-term light curves.}
   {Spectra from 24 objects display spectral features or a high signal-to-noise ratio (S/N). From these, 12 spectroscopic redshifts were determined, ranging from 0.2223 to 0.7018. Furthermore, 1 tentative redshift (0.6622) and 2 redshift lower limits at $z$ > 0.6185 and $z$ > 0.6347 were obtained. The other 9 BL Lacs showed featureless spectra, despite the high S/N ($\geq$ 100) observations. Our comparisons with long-term optical light curves tentatively suggest that redshift measurements are more straightforward during an optical low state of the active galactic nucleus (AGN). Overall, we have determined 37 redshifts and 6 spectroscopic lower limits as part of our programme thus far.}
   {}


\maketitle

\section{Introduction}

Blazars are a type of active galactic nuclei (AGNs) that display exceptional observational properties. Their unique characteristics include beamed non-thermal emission from the radio to high-energy $\gamma$ rays, as well as strongly polarized optical and radio emission ($\ge$ 3\%) \citep[see e.g.][]{Angel80,Angelakis16,Lis11}, variability from a few percent up to a few orders of magnitude on different timescales at all wavelengths \citep[see e.g.][]{Wag95,Falo14}, and (for some) the ejection of superluminal radio blobs \citep[see e.g.][]{VerCo94}. These characteristics are generally explained by Doppler boosting of the jet emission with Lorentz factors of up to $\sim$40 \citep[e.g.][]{jorstad17}.

Recent decades have seen the emergence of a new observational window on the Universe: very-high-energy (VHE; $E$ $>$ 100 GeV) $\gamma$-ray astronomy. Thanks to three major imaging atmospheric Cherenkov telescopes \citep[H.E.S.S.\footnote{\url{https://www.mpi-hd.mpg.de/HESS/}}, MAGIC\footnote{\url{https://magic.mpp.mpg.de/}}, and VERITAS\footnote{\url{https://veritas.sao.arizona.edu/}}; see e.g.][]{HESS, MAGIC1,MAGIC2, VERITAS}, 252 sources have been detected so far at VHE \citep[TeVCAT\footnote{\url{http://tevcat.uchicago.edu/}};][]{Wakely08}, and AGNs make up approximately one-third of these sources. Since 2008, this field has also benefitted from the detection of $\gamma$ rays at high-energy (HE; 100 MeV $<$ $E$ $<$ 100 GeV) with the Large Area Telescope (LAT) on board the {\em Fermi Gamma Ray Space Telescope\footnote{\url{https://fermi.gsfc.nasa.gov/}}} \citep{LAT}. In the coming years, the future world-wide Cherenkov Telescope Array Observatory (CTAO\footnote{\url{https://www.cta-observatory.org/}}) will start its operations with a lower energy threshold for VHE $\gamma$-ray detections, down to a few tens of GeV, and a flux sensitivity improved by a factor of $\sim$10 compared to the current facilities.

Blazars, which include BL Lacs and flat-spectrum radio quasars (FSRQs), are by far the most numerous type of AGNs detected at VHE. The study of their population and evolution is key to high-energy astrophysics (Cherenkov Telescope Array Consortium et al.~2019). However, one of the biggest difficulties in investigating VHE $\gamma$-ray emission from AGNs resides in their limited number, since only 84 of them are known to date. In the future, CTAO is expected to detect hundreds of AGNs of various types and make it possible to carry out population studies. The broad energy range covered by CTAO includes energies where $\gamma$ rays are unaffected by absorption while propagating in the poorly characterized extragalactic background light \citep[EBL; e.g.][]{Gould66, Hauser01, Dwek13} and extends to the VHE regime, where the spectra are strongly distorted by the EBL. This will help reduce systematic effects combining spectra from different instruments, leading to a more reliable EBL determination, thus making it possible to constrain blazar models up to the highest energies with less ambiguity. One challenge in the study of the cosmological evolution of blazars is the difficulty in obtaining redshifts, particularly from the nearly featureless, continuum-dominated spectra of BL Lacs. Indeed, many of the early studies using X-ray or radio-selected samples had highly incomplete redshift measurements, even though the samples were confined to relatively bright sources \citep[BZCAT\footnote{\url{https://www.ssdc.asi.it/bzcat/}};][]{bzcat15}. Uncertainties related to extrapolating unknown redshifts from the measured set of redshifts have further complicated population studies.

The difficulty in measuring BL Lacs redshifts has not yet been solved  and it also plagues  present-day samples of blazars \citep{Shaw13, Pai17a, Mas15a, Mas15b, Pena20}. Indeed, while the redshift distribution of $\gamma$-ray FSRQ ($\ge$ 90 \% of them with known redshifts) displays a maximum at $z$ $\sim$ 1 \citep{Aje12}, the redshift distribution of BL Lacs peaks at $z$ $<$ 0.5 \citep{Aje14}. This is reflected in the luminosity functions of the two classes. The luminosity function of FSRQ has a positive, luminosity-dependent evolution, while BL Lacs have a negative evolution at low luminosities, suggesting a very different evolution among the two classes. However, the results of BL Lacs are very likely biased by the lack of redshifts. For example, considering redshift lower limits from absorbing systems, a relevant population of BL Lacs with redshifts greater than 0.3--0.5 appears plausible. It is thus clear that samples with poor redshift coverage cannot be used to obtain reliable results on the cosmic evolution of $\gamma$-ray BL Lacs \citep{Aje14}. The problem is more severe for VHE BL Lacs, where the number of sources detected at higher redshift becomes sparser (less than ten BL Lacs at $z$ $>$ 0.3, of which two are at $z$ $>$ 0.4). 

The VHE $\gamma$-ray detection of many BL Lacs in different redshift bins at $z$ $>$ 0.3 will allow for a more complete determination of the EBL density (see CTA Consortium et al.~2019). The redshift evolution of diffuse light in the Universe provides a powerful constraint on the galaxy evolution, the formation of stars and galaxies, and cosmology. In this context, the EBL is of fundamental importance. BL Lacs have so far been successfully used to investigate the EBL density \citep[e.g.][]{Aje14, Bit15, Acc19}. However, a major ingredient for its success is a robust redshift determination of the sources used for estimating it. It is therefore of great importance to make an effort to measure the redshift of a large fraction of AGNs detected with {\em Fermi}-LAT and that are likely to be detected with CTAO based on the extrapolation of their LAT spectra.

This is the third paper in a series that aims to determine the redshift of a sample of blazars that are promising targets for CTAO. In the previous two papers, we determined 25 spectroscopic redshifts with values between 0.0838 and 0.8125, 2 tentative redshifts, and 5 lower limits. In this paper, we present detailed results for 24 new targets. The results of the Lick spectra for 17 additional targets in our sample (see Section \ref{sample}) are reported in Appendix \ref{app}.

\section{Sample selection}\label{sample}

In this work, we use the sample introduced in \citet[hereafter Paper~I]{Gol21} and \citet[hereafter Paper~II]{Kasai2023}, which we briefly describe here. The sample has been selected to comprise the best candidates for future observations with the CTAO, extrapolating the spectral information of the hardest sources detected by \emph{Fermi}-LAT at $E$ $>$ 10 GeV to the VHE. 

We started from the sample of BL Lacs and blazar candidates of uncertain type (BCUs) in the Third Fermi LAT Catalog of High energy Sources \citep[3FHL;][]{Fer3FHL17}, where the majority of the 1040 sources (64\%) have no redshift measurement. To estimate their emission in the CTAO band, we extrapolated their average 3FHL energy spectrum, adding a 3 TeV exponential cutoff to simulate the expected spectral curvature in the VHE band. We then performed a Monte Carlo simulation with the Gammapy\footnote{\url{https://gammapy.org}} software \citep{gammapy:2017, gammapy:2019, gammapy:2023}, using the publicly accessible CTAO performance files\footnote{\url{https://www.cta-observatory.org/wp-content/uploads/2019/04/CTA-Performance-prod3b-v2-FITS.tar.gz}}$^{, }$\footnote{\url{https://zenodo.org/record/5163273\#.Yg9-yPVBzPZ}}. To account for the energy- and redshift-dependent $\gamma$-ray opacity of the EBL, we used the model of \citet{Dom11}. For 3FHL sources with no redshift available, we assumed a value of $z = 0.3$, similar to that of $z_{\rm med} = 0.33$ reported by \citet{Shaw13} and $z_{\rm med}$ = 0.285 by \citet{Pena20}. Incidentally, the median redshift of the sources reported in our two previous papers is $z_{\rm med} = 0.255$, which excludes the lower limits with a mean value of $z_{\rm med\,ll} = 0.618$. This value is consistent with this choice. After a first simulation run, we selected a smaller sample on which we performed a literature search that allowed us to correct the redshift of 32 sources (see Paper I) and rerun the simulations. As a result of these simulations, we estimated the minimum observing time required to get a 5$\sigma$ detection with CTAO and the 3FHL sources with no redshift, which require less than 30 hours to be detected, are included in our sample. The final sample contains 165 targets.

\section{Observing strategy}

As in the first two papers in the series (Paper I and Paper II),  our main goal is to determine the spectroscopic redshifts or at least a spectroscopic lower limit for the BL Lacs in our sample. This is achieved by looking at stellar absorption features found in the luminous elliptical galaxies that usually host BL Lacs \citep{Urr00}, such as Ca \textsc{ii} HK doublet, Mg$_b$, and Na \textsc{i} D. Despite the scarcity of their detection in BL Lac spectra, we also searched for emission lines such as [O \textsc{ii}], [O \textsc{iii}], H$\alpha,$ and [N \textsc{ii}]. A spectral resolution, $\lambda$/$\Delta\lambda,$ of a few hundred (ideally up to 1000), along a signal-to-noise ratio (S/N) per pixel of $\sim$100 is required to detect features with EWs $\lesssim$~5~\AA\,expected in these sources. These two conditions have demonstrated good results in the past, as shown in \citet{Pit14} and Papers I and II in our series. In difficult cases, the observations were intended to achieve at least one of these criteria. When possible, observations were carried out during a photometric minimum of the target to benefit from the higher S/N of the host galaxy features in the spectrum due to a reduced non-thermal emission of the central AGN.

\section{Observations and data reduction}

Observations were performed for 41  blazars between 2020-09-20 and 2022-03-26 for a total of 68 hours using the Keck/ESI\footnote{\raggedright Echellette Spectrograph and Imager (ESI) on the Keck II telescope, \url{https://www.keckobservatory.org/about/telescopes-instrumentation}}~\citep{Shei12},  Lick/KAST\footnote{\raggedright{KAST Double Spectrograph on the Shane telescope  at the Lick Observatory}, \url{https://mthamilton.ucolick.org/techdocs/instruments/kast/}}, SALT/RSS\footnote{Robert Stobie Spectrograph (RSS) on the Southern African Large Telescope (SALT), \url{www.salt.ac.za/telescope}}~\citep{Burgh03}, VLT/FORS\footnote{FOcal Reducer and low dispersion Spectrograph (FORS) on the Very Large Telescope (VLT), \url{https://www.eso.org/sci/facilities/paranal/instruments/fors.html}}~\citep{Appenzeller1998}, and GTC/OSIRIS \footnote{Optical System for Imaging and low-Intermediate-Resolution Integrated Spectroscopy (OSIRIS) on the Gran Canarias Telescope (GTC) \url{http://www.gtc.iac.es/instruments/osiris/}}. We note that 
Keck II and GTC have primary mirror diameters of 10 m and SALT's is 11 m, whereas the primary mirror diameter for the VLT is 8.2~m and the mirror diameter of the Shane telescope is 3 m.

 The observations and data analysis follow similar procedures as described in Papers I and II, particularly with respect to the observational configuration, data reduction, flux calibration, telluric correction, and spectral dereddening procedures. After this treatment, the spectral shape is that generally expected for BL Lacs, showing a power-law continuum  superposed with the emission of a host galaxy and (at times) weak emission lines. For some objects with a power law spectrum, namely, WISE J053629.06$-$334302.5, PKS 1424+240, and RBS 1457, as well as (to a smaller extent) 1RXS J203650.9$-$332817 and PMN J2221$-$5224, a flattening is displayed in the red part of the spectrum, which may be due to problems with flat-field frames. In this paper we give only a brief description of the VLT/FORS and GTC/OSIRIS instruments and of their data reduction as data from these instruments were not included in the first two papers in the series.

The visual and near-UV FOcal Reducer and low dispersion Spectrograph
(FORS), part of the VLT of the European Southern Observatory (ESO), is a multi mode (imaging, polarimetry, long slit and multi-object spectroscopy) optical instrument mounted on the VLT UT1 Cassegrain focus \citep{Appenzeller1998}. To achieve the desired coverage, sensitivity, and spectral resolution on most targets, we selected GRISM 600RI+19, given  its sensitivity in the range of 5000--8500 \AA,\ with a spectroscopic resolution of $\lambda$/$\Delta\lambda \sim 800$ and a 40-80\% efficiency. For RBS 1751, we selected GRISM 1200B+97, which is sensitive in the range of 3600--5100 \AA,\ with a spectroscopic resolution of $\lambda$/$\Delta\lambda \sim 1200$ and a 60-80\% efficiency. In all cases, we used a 1.3 arcsec slit, whereas a 5 arcsec slit was used for standard stars. Data reduction was performed using the FORS pipeline \citep{Izz10} up to spectral extraction, which was then performed using Image Reduction and Analysis Facility \citep[IRAF,][]{Tod86}.

OSIRIS is the work-horse imaging and spectroscopic instrument for the GTC \citep{Cep03} installed at the Nasmyth focus. We carried out observations using the R1000B and R1000R gratings, with an overall spectral coverage from 3650 \AA~to 10000 \AA~at a resolution of $\lambda$/$\Delta\lambda$ $\sim$ 1000. The 1.0 arcsec slit was used for the target and for the standard star. We used standard IRAF \citep{Tod86} procedures for long-slit observations to reduce tha data. Table \ref{tabobs1} lists the sources and observational parameters used for the spectroscopic measurements of the 24 targets with spectral features or high S/N featureless spectra. Table \ref{tab_spgraphtech} gives the parameters used in the observations at each telescope. 

\noindent 

All sources, except for 1RXS\,J035000.4+064053 and SUMMS\,J224017$-$524111, have been included not only in the 3FHL, but also in the Second Fermi-LAT Catalog of High-Energy Sources \citep[2FHL;][]{Ackermann16}; therefore, they has been detected by {\em Fermi}-LAT at energies higher than 50 GeV. 1RXS\,J035000.4+064053 and SUMMS\,J224017$-$524111 are the only sources presented in Table \ref{tabobs1} not classified as a BL Lac object in the 4FGL-DR4 catalogue \citep{4FGLDR4}. However, those two sources are classified as high-energy BL Lac objects (HBLs) in the 3HSP catalogue \citep{3HSP}.

Table \ref{tabobs2} in Appendix A presents 26 spectra of 18 blazars observed with Lick/KAST, one of which is also reported in Table \ref{tabobs1} (1RXS J035000.4$+$064053). The 17 remaining are not discussed in detail in this paper because of their low S/N and featureless spectra.



 \begin{sidewaystable*}
\small
\caption{\label{tabobs1} List of observed sources and parameters of the observations for the `main sample' of 24 sources discussed in detail in  Section \ref{s6}. Note: the VLT/FORS2 spectra were all taken with grism GRIS 600RI+19, except for RBS 1751, which was observed with grism GRIS 1200B+97. The sources with a $\dagger$ symbol are listed in the BZCAT catalogue \citep{bzcat15}.} 
\centering
\begin{tabular}{lcccclcclll}
\hline\hline
3FHL name &   4FGL Name & Source name   & RA & Dec  &   Telescope/ &   Slit            & Start Time & Exp.  & Airm. & Seeing      \\
                   &                         &                         &       &         &    Instrument &  (\arcsec)  &  UTC         &  (sec)   &         & (\arcsec)      \\  
 (1)             & (2)                    &             (3)        &  (4) & (5)  &      (6)            &     (7)          &      (8)       &   (9)      &(10)    & (11)           \\ 
\hline
3FHL\,J0051.2$-$6242  & 4FGL\,J0051.2$-$6242   & 1RXS\,J005117.7$-$624154$^{\dagger}$       &  00 51 16.7 & -62 42 04 & VLT/FORS & 1.3 & 2021-10-04 05:53:10 & 1740 & 1.31 & 0.8 \\
3FHL\,J0134.4+2638    & 4FGL\,J0134.5+2637     & 1RXS\,J013427.2+263846$^{\dagger}$         &  01 34 28.2 & +26 38 43 & Keck/ESI   & 1.0 & 2020-12-06 06:45:21 & 6000 & 1.02 & 1.0 \\
3FHL\,J0143.8$-$5846  & 4FGL\,J0143.7$-$5846   & SUMSS\,J014347$-$58455$^{\dagger}$        &  01 43 47.4 & -58 45 51 & SALT/RSS   & 2.0 & 2021-01-09 19:50:38 & 2330 & 1.34 & 1.7 \\
                                      &                                        &                                                                  &                    &                 &                     &       & 2021-01-11 19:31:14 & 2330 & 1.30 & 1.8\\
3FHL\,J0350.0+0640    & 4FGL\,J0350.0+0640     & 1RXS\,J035000.4+064053                           &  03 49 57.8 & +06 41 26 & Keck/ESI   & 1.0 & 2020-12-06 08:36:07 & 7200 & 1.04 & 1.1 \\
3FHL\,J0506.9$-$5434  & 4FGL\,J0506.9$-$5435   & 1ES\,0505$-$546$^{\dagger}$                       &  05 06 57.8 & -54 35 03 & SALT/RSS & 2.0 & 2021-02-06 20:48:22 & 2340 & 1.23 & 1.1 \\
3FHL\,J0536.4$-$3342  & 4FGL\,J0536.4$-$3343   & WISE\,J053629.06$-$334302.5$^{\dagger}$  &  05 36 29.1 & -33 43 03 & VLT/FORS & 1.3 & 2020-10-04 08:07:04 & 1740 & 1.04 & 0.8 \\
3FHL\,J0601.0+3837    & 4FGL\,J0601.0+3838     & B2\,0557+38$^{\dagger}$                               &  06 01 02.9 & +38 38 28 & Keck/ESI & 1.0 & 2020-12-06 11:02:22 & 7100 & 1.07 & 1.1 \\
3FHL\,J0746.3$-$0225  & 4FGL\,J0746.3$-$0225   & WISE\,J074627.03$-$022549.3   &  07 46 27.0 & -02 25 49 & Keck/ESI & 1.0 & 2020-12-06 13:33:45 & 6600 & 1.14 & 0.8 \\
3FHL\,J0826.3$-$6403  & 4FGL\,J0826.4$-$6404   & SUMSS\,J082627$-$640414                    &  08 26 27.1 & -64 04 21 & VLT/FORS & 1.3 & 2021-12-01 06:58:20 & 2610 & 1.34 & 0.6  \\
3FHL\,J0858.0$-$3131  & 4FGL\,J0858.0$-$3130 &  1RXS\,J085802.6$-$313043                   &  08 58 02.9 & -31 30 38 & VLT/FORS & 1.3 & 2022-01-02 04:49:38 & 2610 & 1.09 & 0.8 \\
                                    &                                        &                                                                   &                   &                 &                    &       & 2022-02-12 04:13:32 & 2610 & 1.01 & 1.1 \\
                                    &                                        &                                                                   &                   &                 &                    &       & 2022-03-04 02:08:00 & 2610 & 1.01 & 0.6 \\
3FHL\,J1130.5$-$7801  & 4FGL\,J1130.5$-$7801  & SUMSS\,J113032$-$780105                   &  11 30 31.9 & -78 01 05 & VLT/FORS & 1.3 & 2021-01-31 07:43:24 & 2600 & 1.72 & 1.2 \\
                                     &                                        &                                                                   &                    &                 &                   &       & 2022-03-26 02:14:48 & 2190 & 1.71 & 0.7\\
3FHL\,J1427.0$+$2348 & 4FGL\,J1427.0+2348 & PKS\,1424+240$^{\dagger}$ & 14 27 00.4 & +23 48 00 & GTC/OSIRIS & 1.0 & 2021-04-22 05:22:11 & 1800 & 1.67 & 0.9 \\
3FHL\,J1445.0$-$0326   &  4FGL\,J1445.0$-$0326  &  RBS\,1424$^{\dagger}$                                 &  14 45 06.2 & -03 26 13 & SALT/RSS & 2.0 & 2021-04-07 01:35:38. & 2390 & 1.26 & 0.9 \\
3FHL\,J1503.7$-$1540   &  4FGL\,J1503.7$-$1540  &  RBS\,1457$^{\dagger}$                                 &  15 03 40.7 & -15 41 14 & VLT/FORS & 1.3 & 2021-04-15 06:07:40 & 1950 & 1.02 & 0.8 \\
3FHL\,J1539.7$-$1127   &  4FGL\,J1539.7$-$1127  &  PMN\,J1539$-$1128                                  &  15 39 41.2 & -11 28 35 & SALT/RSS & 2.0 & 2021-04-23 02:21:26 & 2190 & 1.33 & 1.0 \\
                                      &                                        &                                                                   &                    &                 &                     &       & 2021-04-24 01:56:38 & 2190 & 1.26 & 1.2\\
3FHL\,J1656.9$-$2010   &  4FGL\,J1656.9$-$2010  &  1RXS\,J165655.0$-$201049                      &  16 56 55.1 & -20 10 56 & VLT/FORS & 1.3 & 2021-04-18 05:49:59 & 2600 & 1.10 & 1.0 \\
3FHL\,J1745.4$-$0752  &  4FGL\,J1745.4$-$0753  &  TXS\,1742$-$078$^{\dagger}$                   &  17 45 27.1 & -07 53 04 & VLT/FORS & 1.3 & 2021-04-15 07:20:17 & 2600 & 1.10 & 1. 1\\
                                     &                                        &                                                                    &                    &                 &                    &       & 2021-04-22 08:47:38 & 2600 & 1.07 & 0.8\\
3FHL\,J1842.4$-$5841  &  4FGL\,J1842.4$-$5840  &  1RXS\,J184230.6$-$584202                       & 18 42 29.8 & -58 41 56 & VLT/FORS & 1.3 & 2020-09-20 01:03:25  & 1700 & 1.29 & 0.8 \\
                                     &                                         &                                                                  &                   &                &                    &        & 2021-04-15 08:23:25 & 2400 & 1.24 & 1.4 \\
3FHL\,J2036.9$-$3328 &  4FGL\,J2036.9$-$3329   &  1RXS\,J203650.9$-$332817                      &  20 36 49.5 & -33 28 31 & VLT/FORS & 1.3 & 2021-04-23 08:08:38 & 3600 & 1.20 & 1.1 \\
                                     &                                         &                                                                   &                   &                &                     &    & 2021-05-09 07:36:22  & 3600 & 1.12 & 0.8 \\
3FHL\,J2131.0$-$2746 &  4FGL\,J2131.0$-$2746   &  RBS\,1751$^{\dagger}$                               &  21 31 03.2 & -27 46 58 & VLT/FORS & 1.3 & 2020-09-21 05:41:55 & 350 & 1.04 & 0.8 \\
3FHL\,J2156.0+1818    &  4FGL\,J2156.0+1818      &  RX\,J2156.0+1818                                      &  21 56 01.6 & +18 18 37 & Keck/ESI & 1.0 & 2020-12-06 04:49:01 & 6000 & 1.17 & 1.0 \\
3FHL\,J2221.5$-$5226 &  4FGL\,J2221.5$-$5225   &  PMN\,J2221$-$5224$^{\dagger}$              &  22 21 29.3 & -52 25 28 & VLT/FORS & 1.3 & 2020-09-21 05:50:31 & 900 & 1.40 & 0.7 \\
3FHL\,J2240.3$-$5240 &  4FGL\,J2240.3$-$5241   &  SUMSS\,J224017$-$524111                      &  22 40 17.6 & -52 41 13 & VLT/FORS & 1.3 & 2020-09-21 05:41:55 & 2600 & 1.43 & 0.7 \\
3FHL\,J2243.7$-$1232 &  4FGL\,J2243.7$-$1231   &  RBS\,1888                                                &  22 43 41.6 & -12 31 38 & VLT/FORS & 1.3 & 2020-09-22 02:12:36 & 3900 & 1.43 & 0.8 \\

\hline
\hline

\end{tabular}
\tablefoot{The columns contain:  (1) 3FHL Name, (2) 4FGL Name, (3) Source Name, a $^{\dagger}$ indicates the source is in BZCat, (4) Right Ascension (J2000), (5) Declination (J2000), (6) Telescope and Instrument, (7) Slit Width in arcsec, (8) Start Time of the observations, (9) Exposure Time, (10) Average Airmass, and (11) Average Seeing.}
\end{sidewaystable*}

\begin{table*}
\caption{\label{tab_spgraphtech} Spectroscopic mode, wavelength coverage, and throughput and spectral resolution of the five spectrographs, as used in this work.
}

\centering
\begin{tabular}{lcccc}
\hline\hline

Instrument Name & Spectroscopic mode &  Wavelength coverage (\AA) & Throughput $p$ & Spectral resolution $\lambda$ / $\Delta \lambda$ \\

\hline

Keck/ESI  & Echellette & 3900 - 10000 & $p \geq$ 28\% & $\sim$ 10000 \\
Lick/KAST & Blue arm & 3500 - 5600 & 5\% < $p$ < 30\%  & $\sim$ 1000 \\
Lick/KAST & Red arm & 5400 - 8000 & 30\% < $p$ < 40\%  & $\sim$ 1500 \\

SALT/RSS & Long slit & 4500 - 7500  & $p$ > 20\% & $\sim$ 1000 \\
VLT/FORS & Low resolution, GRISM 600RI+19 & 5000 - 8500  & 20\% < $p$ < 30\% & $\sim$ 800 \\
VLT/FORS & Low resolution, GRISM 1200B+97 & 3600 - 5100  & 15\% < $p$ < 20\% & $\sim$ 1200 \\
GTC/OSIRIS & Long slit, GRISM R1000B & 3650 - 7500  & 15\% < $p$ < 20\% & $\sim$ 1000 \\
GTC/OSIRIS & Long slit, GRISM R1000R & 5100 - 10000  & 15\% < $p$ <20\% & $\sim$ 1000 \\

\hline
\end{tabular}
\end{table*}

\section{Redshift measurement and estimation of the blazar total emission}

To determine the redshift, we searched for absorption and emission features in the spectra, requiring a minimum of two independent detections corresponding to the same redshift for a robust determination. Details on the lines we searched can be found in Paper II. For each feature, we measured the EW by fitting the continuum with cubic splines and integrating the flux for each pixel. The uncertainties in the EW are determined by the quadrature sum of the normalized flux errors and the continuum placement contribution \citep[see][]{Sem92}.

The redshift uncertainty is estimated by considering the uncertainty in the position of the detected features. The feature position and its uncertainty were determined by a Gaussian fit. To this uncertainty, we added in quadrature the wavelength calibration uncertainty. The results are presented in Tables \ref{tabeqw}, \ref{tabeqw2}, \ref{SUMSStab}, and \ref{tabres1}. In the case of SUMMS\,J224017$-$524111, by considering the large number of emission lines detected, the results are reported in a dedicated table.

We modelled the SED with the sum of a power law for describing the jet emission and an elliptical galaxy template for describing the host \citep{Man01, Bru03}. When needed, Gaussian emission features are added in the modelling  \citep{Pit14} and only one template was used per spectrum for simplicity. Only two free parameters are needed: the power -law slope and jet-to-galaxy ratio. Table \ref{tabres1} shows the results of the fits. 

We also estimated the absolute magnitude of the detected host galaxies. Slit losses are estimated by assuming the host galaxy effective radius, $r_{\rm e}$, to be 10 kpc for a de Vaucouleurs profile. From the template spectra, we computed the K-corrections and did not apply evolutionary corrections. In the case of non-detection of a host, the spectra are fitted with a power law and normalised at the band centre. Due to the high S/N of the spectra, the relative errors on the fitted parameters are very small (10$^{-3}$ or so). However, there is residual curvature seen in the spectra that maybe intrinsic (i.e. slope change) or due to calibration issues (fluxing and/or extinction correction). To take this into account, we fit  the red and blue halves of the spectra separately and take the difference between the blue and the red parameters as 1 sigma errors on the parameters of the total spectrum. The uncertainties quoted in Table \ref{tabres1} are three times (i.e. three sigma) these values.

\begin{table*}
\caption{\label{tabeqw} Equivalent widths in \AA~of the absorption features detected in the spectra at the measured redshift for each source. }
\centering
\begin{tabular}{lccccc}
\hline\hline

Source {name} &  CaHK & CaIG & Mgb & CaFe & NaID  \\
            &       &      &     &       &     \\
 (1)  &  (2)   &  (3) & (4) & (5) &  (6)  \\        

\hline

SUMSS\,J014347$-$58455             & 1.3 $\pm$ 0.1          & 1.2 $\pm$ 0.2  &   ---                   & ---                     &  ---                  \\
1RXS\,J035000.4$+$064053           & 6.6 $\pm$ 0.7           & 4.1 $\pm$ 0.4 & 7.3 $\pm$ 0.5   & 1.3 $\pm$ 0.4   &  4.0 $\pm$ 0.3   \\
B2\,0557$+$38                      & 0.8 $\pm$ 0.7$^*$   & ---                 &   ---                    & ---                     & --- \\
SUMSS\,J082627$-$640414            & 2.0 $\pm$ 0.4         & 2.0 $\pm$ 0.4  &  1.4 $\pm$ 0.2   & 0.3 $\pm$ 0.1    & 2.7 $\pm$ 0.4  \\
1RXS\,J085802.6$-$313043           & 4.7 $\pm$ 0.7        & 1.2 $\pm$ 0.3  &   ---                   & ---                     &  ---                  \\
SUMSS\,J113032$-$780105            & 2.0 $\pm$ 0.5          & 1.0 $\pm$ 0.2 &  3.0 $\pm$ 0.3   & 1.0 $\pm$ 0.2     & 2.2 $\pm$ 0.2  \\
PMN\,J1539$-$1128                  & 1.2 $\pm$ 0.2          &  ---                 & ---                     &    ---                &     ---               \\
1RXS\,J165655.0$-$201049           & 2.1 $\pm$ 0.4         &   ---     &  0.6 $\pm$ 0.1  &   0.4 $\pm$ 0.2      &0.9 $\pm$ 0.2  \\
1RXS\,J184230.6$-$584202           & 3.2 $\pm$ 0.5         & 0.6 $\pm$ 0.2  &  1.5 $\pm$ 0.4   & 0.8 $\pm$ 0.2      &   --- \\
SUMMS\,J224017$-$524111            &   ---                       & 1.6 $\pm$ 0.2  &  3.2 $\pm$ 0.2    & 1.0 $\pm$ 0.2      & 1.8 $\pm$ 0.2 \\
RBS\,1888                          &    ---                      &  3.5 $\pm$ 0.3 &  4.5 $\pm$ 0.2    & 2.3 $\pm$ 0.2      &  2.2 $\pm$ 0.2 \\
\hline
\hline
\end{tabular}
\tablefoot{The CaIG feature of SUMSS\,J082627$-$640414 is likely contaminated by the $\lambda$ 5780 and  $\lambda$ 5796 DIBs. The CaIG feature of 1RXS\,J085802.6$-$313043 is likely contaminated by telluric H2O.
The columns are (1) Source name, (2) Equivalent width of the CaHK feature with errors, (3) Equivalent width of the CaIG feature with errors, (4) Equivalent width of the Mgb feature with errors, (5) Equivalent width of the CaFe feature with errors, (6) Equivalent width of the NaID feature with errors. If the feature is not detected, the legend is `---'. The detection of CaHK in B2 0557+38 is uncertain and it is flagged with an asterisk.}
\end{table*}

\begin{table}
\small
\caption{\label{tabeqw2} Equivalent width in \AA~of the main emission features detected in the spectra at the measured redshift. The emission features detected in the spectrum of SUMMSJ224017$-$5241 are in Table \ref{SUMSStab}.}

\centering
\begin{tabular}{lccc}
\hline\hline
Source name &      [OII]     &      [OIII]                 &     [OIII]      \\
                      &  $\lambda$ 3727                &    $\lambda$ 4959    &   $\lambda$ 5007                   \\
               (1)  &   (2)          &   (3)                          &   (4)     \\      
\hline
B2\,0557$+$38                  &  0.6 $\pm$ 0.2    & ---                  & ---               \\
PKS\,1424$+$240   &  0.15 $\pm$ 0.02              & 0.10 $\pm$ 0.02  & 0.35 $\pm$ 0.02 \\
PMN\,J1539$-$1128          &  0.4 $\pm$ 0.03  & ---                  &   ---               \\
TXS\,1742$-$078              & 1.7 $\pm$ 0.4                    & 0.8 $\pm$ 0.2  & 3.0 $\pm$ 0.2  \\
\hline
\hline
\end{tabular}
\tablefoot{Columns are: (1) Source name, (2) Equivalent width of the [OII]$\lambda$ 3727 feature with errors, (3) Equivalent width of the [OIII]$\lambda$ 4959 feature with errors, (4) Equivalent width of the [OIII]$\lambda$ 5007   with errors. If the feature is not detected, the legend is  `---'.}
\end{table}

\begin{table}
\caption{ Equivalent widths (in \AA)~of the emission lines detected in SUMMS\,J224017$-$524111.}
\begin{center}
\begin{tabular}{cc}
\hline
\hline
  Line &  Equivalent width \\
          &     \AA     \\
   (1) &    (2)  \\          
\hline
 {[OIII]$\lambda$ 4959}    &   0.3 $\pm$ 0.1   \\
 {[OIII]$\lambda$ 5007}    &   0.7 $\pm$ 0.1   \\
 {[OI]$\lambda$ 6300}      &   0.8 $\pm$ 0.1   \\
 H$\alpha$                 &   1.7 $\pm$ 0.1   \\
 {[NII]$\lambda$ 6548}     &   0.8 $\pm$ 0.2   \\
 {[NII]$\lambda$ 6583}     &   2.3 $\pm$ 0.1   \\
 {[SII]$\lambda$ 6716}     &   0.6 $\pm$ 0.1   \\
 {[SII]$\lambda$ 6731}     &   0.6 $\pm$ 0.1   \\
\hline
\hline

\end{tabular}
\end{center}
\label{SUMSStab}
\end{table}%

\section{Sources and results}\label{s6}

For the 24 sources in Table \ref{tabobs1}, twelve spectroscopic redshifts 
were determined, ranging from 0.2223 to 0.7018, one tentative redshift and two redshift lower limits ($z$ $>$ 0.6185 and $z$ $>$ 0.6347) were obtained. In the following we discuss the results of our observations for each of the sources.

\begin{figure*}
   \centering
 \includegraphics[width=6.8truecm,height=5.75truecm]{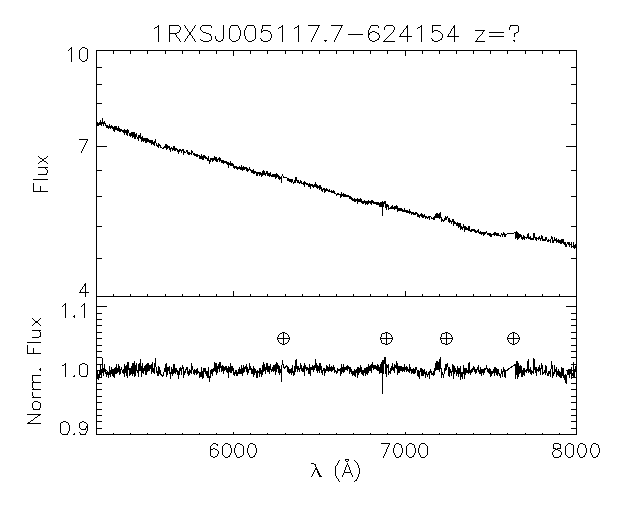}  \includegraphics[width=6.8truecm,height=5.75truecm]{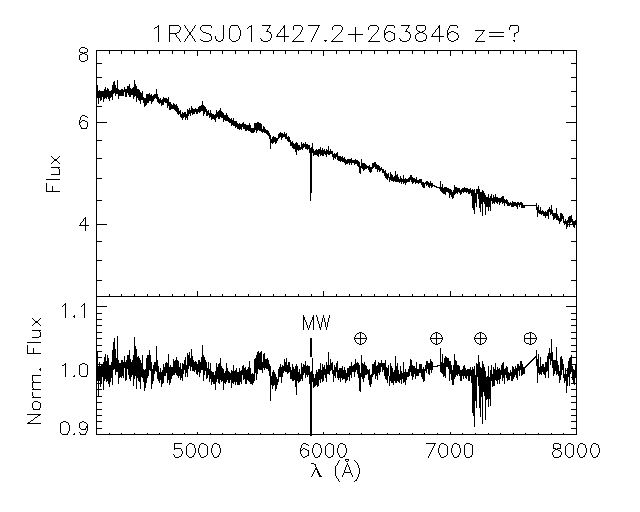} 
 \includegraphics[width=6.8truecm,height=5.75truecm]{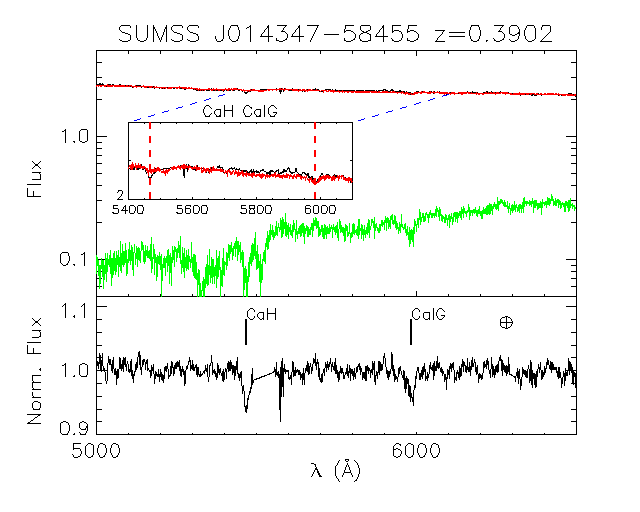}  \includegraphics[width=6.8truecm,height=5.75truecm]{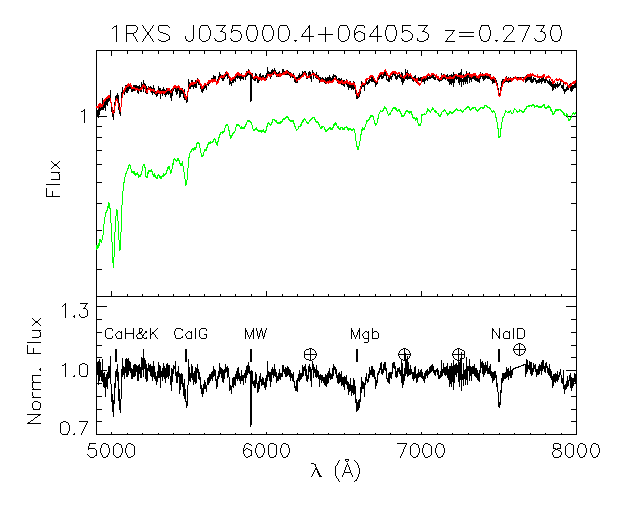}
  \includegraphics[width=6.8truecm,height=5.75truecm]{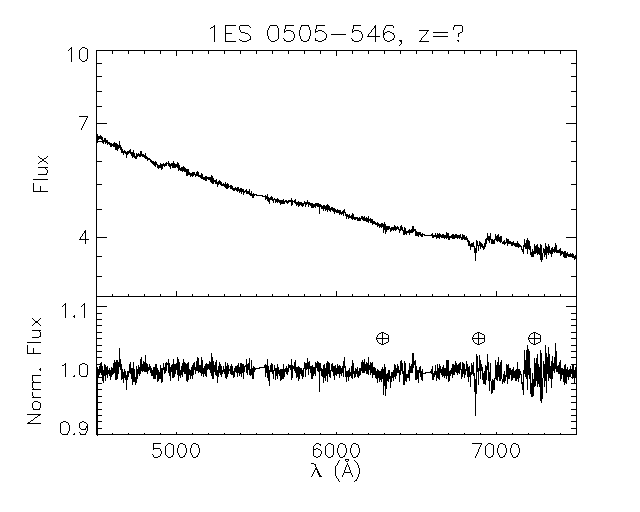}  \includegraphics[width=6.8truecm,height=5.75truecm]{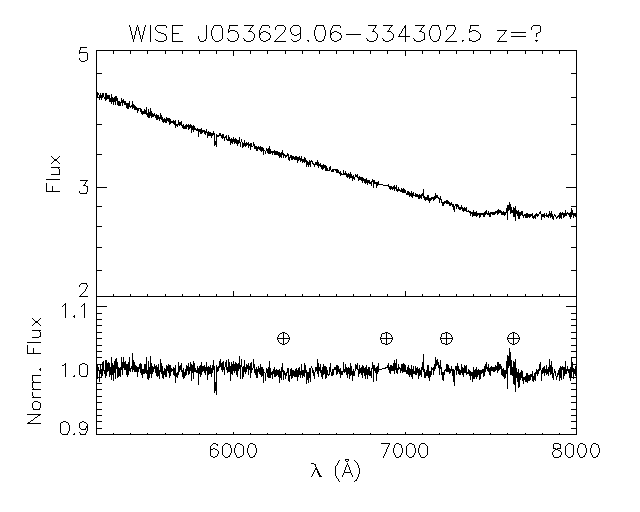}
 \includegraphics[width=6.8truecm,height=5.75truecm]{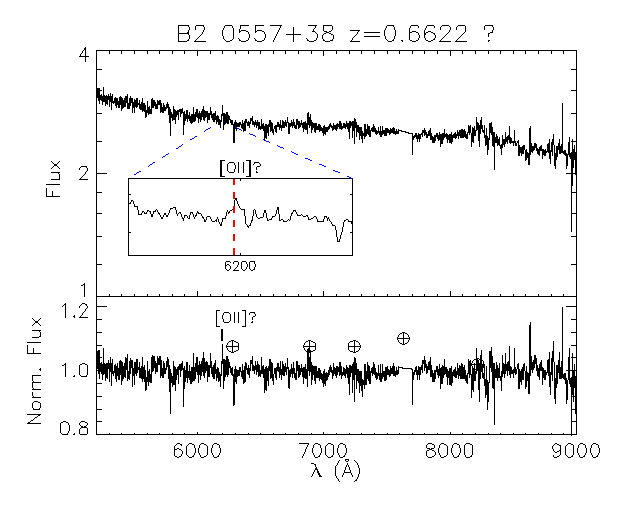}  \includegraphics[width=6.8truecm,height=5.75truecm]{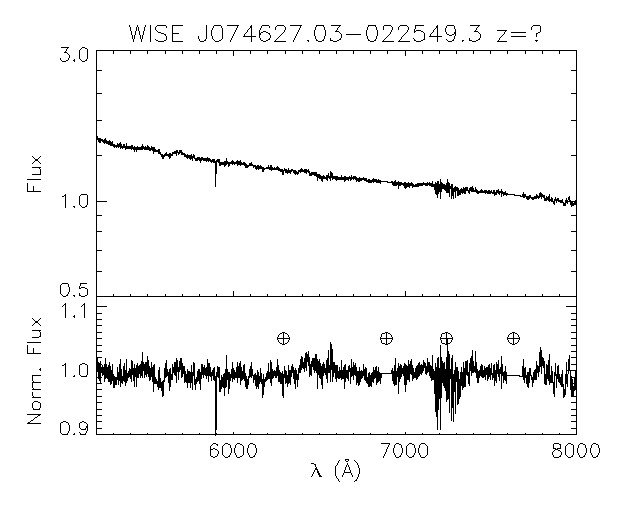}
   \caption{
   Flux-calibrated and normalized spectra of the first eight sources in Table \ref{tabobs1}. Each panel contains the spectrum, continuum, and galaxy model for a given source.  Each panel has  two parts. Upper:  Flux-calibrated and telluric-corrected spectrum (black) alongside the best fit model (red). The flux is in units of 10$^{-16}$ erg  cm$^{-2}$ s$^{-1} $\AA$^{-1}$. The elliptical galaxy component is shown in green. Lower:  Normalised spectrum with labels for the detected absorption features. Atmospheric telluric absorption features are indicated by the symbol $\oplus$ and Galactic absorption features are labelled  `MW'. }

\label{fig_spec1}
    \end{figure*}

\subsection{1RXS J005117.7\texorpdfstring{$-$}{-}624154 }

A moderate S/N featureless EFOSC spectrum has been reported by \citet{Mase13}, while Marais \& van Soelen (in prep.) suggest a possible redshift of $z$ = 0.156. We observed the source with VLT/FORS for an exposure time of 1740 sec, but despite a very high S/N (212), the obtained spectrum  is featureless (see Fig.~\ref{fig_spec1}, top panel, on the left).

\subsection{1RXS J013427.2\texorpdfstring{$+$}{+}263846 }

A low S/N featureless spectrum obtained with the Hobby–Eberly Telescope (HET) has been reported by \citet{Shaw13}. A moderate S/N Mayall/KOSMOS spectrum reported by \citet{Marchesi18} shows the detection of a bright emission line. If it is interpreted as MgII, this line indicates $z$ = 0.571. We observed the source with Keck/ESI for an exposure time of 6000 sec. The resulting high S/N (135) spectrum (see Fig.~\ref{fig_spec1}, top panel, on the right) is featureless. The flux estimated at 5000 \AA\, in our spectrum is only a factor of 1.5 higher than the flux in the spectrum reported in \citet{Marchesi18}, suggesting that the activity state of the AGN at the time of the two observations is not so different. We suspect that the MgII emission line reported in \citet{Marchesi18} is an artefact.

\subsection{SUMSS J014347\texorpdfstring{$-$}{-}58455   }

Previous spectra obtained by SOAR, with a moderate S/N \citep{Lan15}, and by EFOSC, with a low S/N \citep{Titov17}, are featureless. We performed two separate observations with SALT/RSS at distance of two days with a total exposure time of 4660 sec. The total averaged spectrum with a S/N of 113 is presented in Fig.~\ref{fig_spec1} (second row panel, on the left). We detected the CaH and CaIG, providing a redshift $z$ = 0.3902 $\pm$ 0.0001. With SALT, the CaK line falls into a CCD gap.

\subsection{1RXS J035000.4\texorpdfstring{$+$}{+}064053 }

No previous spectra are available in literature. We observed this source with Keck/ESI for an exposure time of 7000 sec with a S/N of 46. We clearly detect in the spectrum (Fig.~\ref{fig_spec1}, second row panel, on the right) the host galaxy with several absorption features at $z$ = 0.2730 $\pm$ 0.0002. We note that the jet-to-galaxy flux ratio is low at 0.4 $\pm$ 0.1, resulting in quite intense absorption lines, allowing for their detection even in a relatively low S/N spectrum. The host galaxy is quite luminous at $M_{R}$ = $-$23.1. 

\subsection{1ES 0505\texorpdfstring{$-$}{-}546}

A featureless EFOSC spectrum with intermediate S/N has been reported by \citet{Mase13}. We observed the source with SALT/RSS for an exposure time of 2340 sec. Our SALT/RSS spectrum (Fig.~\ref{fig_spec1}, third row panel, on the left) has a high S/N ($\sim$120) and it is featureless.

\begin{table*}
\caption{\label{tabres1} Analysis results for the sources of the `main sample'. As the redshift of B2\,0557$+$38 is uncertain, for this source we also present the results of a simple power-law fit. The spectral bin width is 1\,\AA~for sources observed with Keck/ESI, GTC/OSIRIS, and SALT/RSS and 1.66\,\AA~for sources observed with FORS.}
\centering
\begin{tabular}{lccccccc}
\hline\hline

 Source name  & S/N &  R$_{\rm c}$(BL Lac) & Redshift   & Flux ratio  & R$_{\rm c}$(gal) & M$_{\rm R}$ & Slope   \\
              &     &   (obs)              &            &            &  (fit)            &   (gal)     &         \\  
   (1)  & (2) & (3)    &  (4)  &  (5)   &  (6)      &  (7)   &  (8) \\      
\hline 
1RXS\,J005117.7$-$624154         & 212   & 16.3 $\pm$ 0.1    &   ---                               &  ---               &  ---                      &   ---      & -1.1 $\pm$ 0.1  \\ 
1RXS\,J013427.2$+$263846        &  135 &  16.7 $\pm$ 0.2   &    ---                               &  ---                &  ---                     &  ---       & -1.1 $\pm$ 0.2  \\
SUMSS\,J014347$-$58455           &   113   &    17.4 $\pm$ 0.1        &    0.3902 $\pm$ 0.0001      &     6.1 $\pm$ 1.7             &  19.5 $\pm$ 0.3                &    -22.6        &    -1.1 $\pm$ 0.1                  \\
1RXS\,J035000.4$+$064053        &    46    &    17.4 $\pm$ 0.1                       &  0.2730 $\pm$ 0.0002 &  0.4 $\pm$ 0.1        &   17.9 $\pm$ 0.1       &    -23.1        &     -1.6 $\pm$ 0.1                    \\ 
1ES\,0505$-$546                           &  126   & 16.8 $\pm$ 0.1   &    ---                                &  ---                &  ---                      &   ---     & -1.2 $\pm$ 0.3                      \\      
WISE\,J053629.06$-$334302.5     & 172   & 17.1 $\pm$ 0.1 &    ---                                 & ---                 &   ---                     &   ---     &  -1.2 $\pm$ 0.2   \\
B2\,0557$+$38                                &    38   &    16.9 $\pm$ 0.1                    & 0.6622$?$ &   5.7 $\pm$ 0.1                   &    19.9 $\pm$ 0.1                         &    -24.3       &   -1.0 $\pm$ 0.3                     \\
B2\,0557$+$38                                &    38   &  16.9 $\pm$ 0.1                        & --- &    ---                  &     ---                        &   ---        &    -0.4 $\pm$ 0.1                    \\
WISE\,J074627.03$-$022549.3      & 112   & 18.2 $\pm$ 0.2 &  ---                                   &    ---                  &    ---                         &      ---      & -1.0 $\pm$ 0.2    \\
SUMSS\,J082627$-$640414          &   112   &   17.6 $\pm$ 0.1             & 0.3397 $\pm$ 0.0002       & 4.1 $\pm$ 0.5                     &       19.3 $\pm$ 0.2                      &    -22.3        &      -1.5 $\pm$ 0.1                 \\
1RXS\,J085802.6$-$313043            &   68    &     19.3 $\pm$ 0.1                    &  0.7018 $\pm$ 0.0002       & 3.3 $\pm$ 0.5                      &  21.5 $\pm$ 0.2                           &    -23.0        &     -1.5 $\pm$ 0.4                  \\   
SUMSS\,J1130.5$-$780105             &  119    &  17.1 $\pm$ 0.1                      & 0.3171 $\pm$ 0.0002        &        3.2 $\pm$ 0.1               &        18.2 $\pm$ 0.2                     &   -23.2         &     -1.2 $\pm$ 0.1                   \\
PKS\,1424$+$240                      &  305   & 14.5 $\pm$ 0.1 & 0.6045 $\pm$ 0.0002 &    ---  &   --- & --- & -0.9 $\pm$ 0.1 \\
RBS\,1424                                     &  134  & 16.9 $\pm$ 0.1 &  ---                                   &  ---                   &  ---                       & ---       & -1.1 $\pm$ 0.1   \\
RBS\,1457                                     & 119   & 17.4 $\pm$ 0.2 &   ---                                  &    ---                       &   ---                          &   ---         & -1.3 $\pm$ 0.1    \\
PMN\,J1539$-$1128                      &    154  &   17.9 $\pm$ 0.1                       &  0.4420 $\pm$ 0.0002     &   6.2 $\pm$ 0.8                       &    19.0 $\pm$ 0.3                         & -23.5   & -1.2$\pm$0.2     \\
1RXS\,J165655.0$-$201049        & 131 &     16.9 $\pm$ 0.1                     &  0.3405 $\pm$ 0.0002     &     9.0 $\pm$ 2.5                     &        18.6 $\pm$ 0.3                      &     -23.0    &     -1.0 $\pm$ 0.1                      \\
TXS\,1742$-$078                          & 50   & 17.8 $\pm$ 0.1   &  0.4210 $\pm$ 0.0002      &         ---          &          ---            &    ---      & -0.6 $\pm$ 0.1   \\
1RXS\,J184230.6$-$584202          & 77  &   17.8 $\pm$ 0.1                & 0.4226 $\pm$ 0.0003     &   5.6 $\pm$ 0.1       &     19.4 $\pm$ 0.1                        &    -22.9      &   -1.3 $\pm$ 0.1                     \\ 
1RXS\,J203650.9$-$332817         & 143  &  18.1 $\pm$ 0.1  &  ---                                 &  ---                     &  ---                       &  ----   & -1.1 $\pm$ 0.1 \\
RBS\,1751                                     &   23 & 18.3 $\pm$ 0.1$^{\dagger}$    & $\ge$ 0.6185 $\pm$ 0.0001   &  ---         &  ---                       &  ---     & -1.2 $\pm$ 0.2    \\
RX\,J2156.0$+$1818                    &  95  &    17.5 $\pm$ 0.2                    & $\ge$ 0.6347 $\pm$ 0.0001                      & ---          & ---                        &  ---     &  -1.6 $\pm$ 0.1   \\
PMN\,J2221$-$5224                     & 151   & 16.5 $\pm$ 0.1 &  ---                                                  &   ---                       &     ---              &      ---        &   -1.0 $\pm$ 0.1 \\
SUMMS\,J224017$-$5241            &    94     &    17.6 $\pm$ 0.1                    &  0.2223 $\pm$ 0.0001                  &    2.6 $\pm$ 0.6                    &   18.3 $\pm$ 0.2                &    -22.1    &             -0.5 $\pm$ 0.1       \\
RBS\,1888                                    &   63  &  18.6 $\pm$ 0.1                      &  0.2265$\pm$0.0002                  &   0.2 $\pm$ 0.1                      &     18.2 $\pm$ 0.2              &     -22.3        &    2.0 $\pm$ 0.6                        \\
\hline
\hline
\end{tabular}
\tablefoot{Columns are (1) Source name; (2) Median S/N per spectral bin measured in continuum regions; (3) R$_{\rm c}$, Cousins magnitude of the BL Lac spectrum corrected for reddening, telluric absorption, and slit losses with errors. Slit losses were estimated using an effective radius r$_e$=10 kpc for all sources; (4) Redshift or lower limit with error, $?$ indicates a tentative redshift, (5) Flux Ratio jet/galaxy at 5500 \AA~in rest frame; (6) R$_{\rm c}$, Cousins magnitude of the galaxy with the same corrections as in column (3); (7) Absolute R magnitude of the galaxy, the errors are the same as those in  column (6); (8) Power-law slope with errors.  If the entry is unknown, the legend is `---'.

$^{\dagger}$ B Magnitude

$?$ Uncertain redshift

}
\end{table*}

\subsection{WISE J053629.06\texorpdfstring{$-$}{-}334302.5  }
  
 This BL Lac has been observed with NTT/EFOSC by \citet{Shaw13}, obtaining a medium S/N ($\sim$ 40) featureless spectrum. We observed it with VLT/FORS for 1740 s. The resulting spectrum with S/N $\sim$ 170 (Fig.~\ref{fig_spec1}, third row panel, on the right) has no detectable spectral features. The redshift of WISE\,J053629.06$-$334302.5 remains unknown.
  
\subsection{B2 0557\texorpdfstring{$+$}{+}38  }

B2\,0557+38 is a rather weak (i(PanStarrs)=18.4) absorbed ($E_{B-V}$ = 0.484) source in the optical range. A low S/N featureless MMT spectrum has been presented in \citet{Pag14}. The source was observed also by \citet{Pai20} using GTC/OSIRIS with a moderate S/N spectrum. In that spectrum, they tentatively detect the Ca II doublet at $z$ = 0.662. We observed the source with Keck/ESI for 7200 s. In our Keck spectrum (Fig.~\ref{fig_spec1}, bottom panel, on the left) the detection of Ca II HK absorption feature is uncertain, but we possibly detect [OII] at $z$ = 0.6622. Given the low S/N of the feature (38), we consider this as a tentative detection. The redshift of B2\,0557+38 remains tentative. The host galaxy is very luminous at $M_{R}$ = $-$24.3.

\subsection{WISE J074627.03\texorpdfstring{$-$}{-}022549.3}

An intermediate S/N featureless spectrum taken with NOT with an integration time of 1200 s has been reported by \citet{Marchesini16}. We performed an observation with Keck/ESI for 6600 s, obtaining a high S/N ($\sim$ 110) spectrum (Fig.~\ref{fig_spec1}, bottom panel, on the right), but it is featureless. The redshift of WISE\,J074627.03$-$022549.3 remains unknown.


\subsection{SUMSS J082627\texorpdfstring{$-$}{-}640414 }

No previous spectrum of the source is available in literature. We performed a 2610 s observation with VLT/FORS obtaining a S/N = 112 spectrum  (Fig.~\ref{fig_spec2}, top panel, on the left), and we are able to detected Ca II HK and other stellar features (CaIG, Mg$_{b}$, CaFe, NaID) obtaining a precise redshift of $z$ = 0.3397 $\pm$ 0.0002.

\subsection{1RXS J085802.6\texorpdfstring{$-$}{-}313043   }

 A moderate S/N featureless spectrum obtained at SALT with Goodman High Throughput spectrograph for 1050 s has been presented in \citet{Pen17}. We performed three observations with VLT/FORS for a total exposure time of 7830 s, obtaining a S/N = 68 spectrum (Fig.~\ref{fig_spec2}, top panel, on the right). We are able to detect Ca II HK and Ca I G at a redshift $z$ = 0.7018 $\pm$ 0.0002.


\subsection{SUMSS J113032\texorpdfstring{$-$}{-}780105}

\begin{figure*}[htbp!]
   \centering
 \includegraphics[width=6.8truecm,height=5.75truecm]{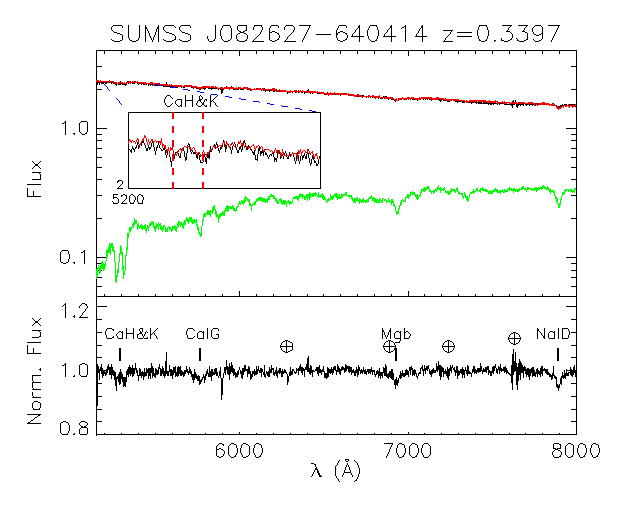} \includegraphics[width=6.8truecm,height=5.75truecm]{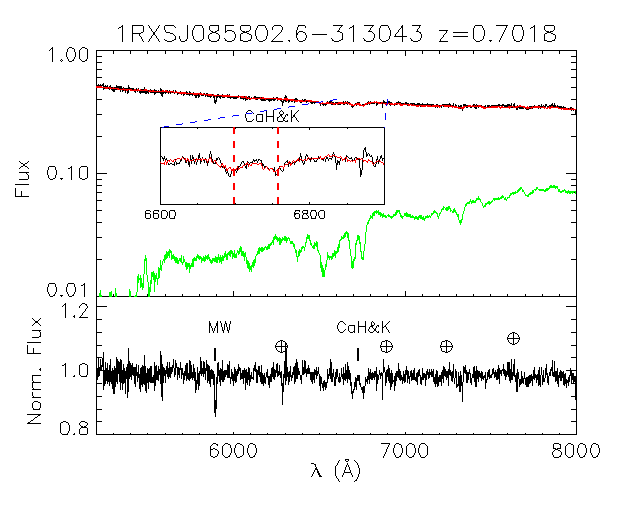}  
  \includegraphics[width=6.8truecm,height=5.75truecm]{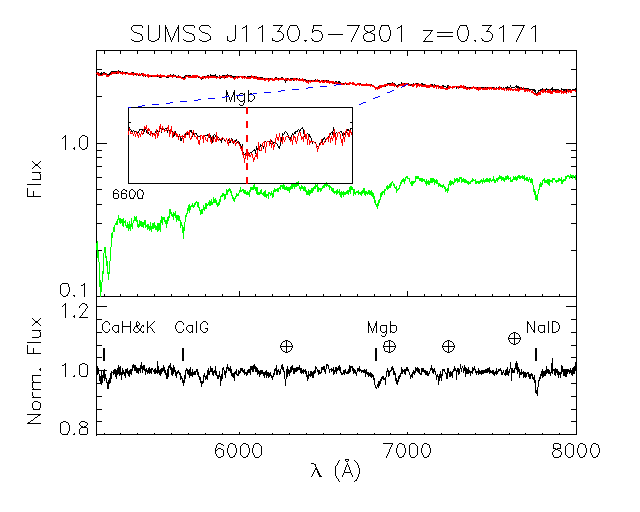} \includegraphics[width=6.8truecm,height=5.75truecm]{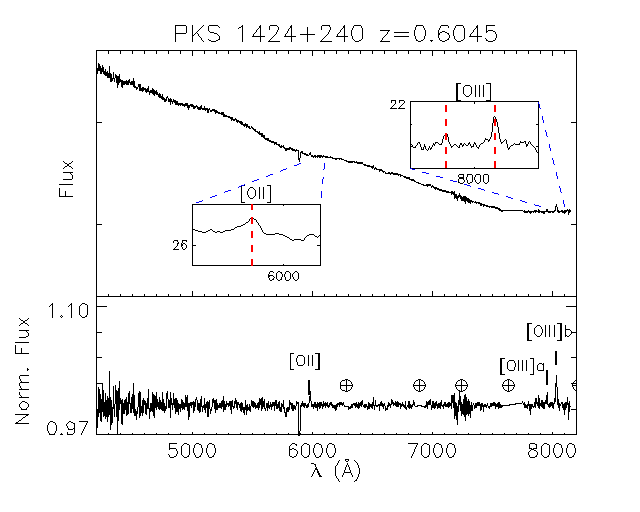}  \includegraphics[width=6.8truecm,height=5.75truecm]{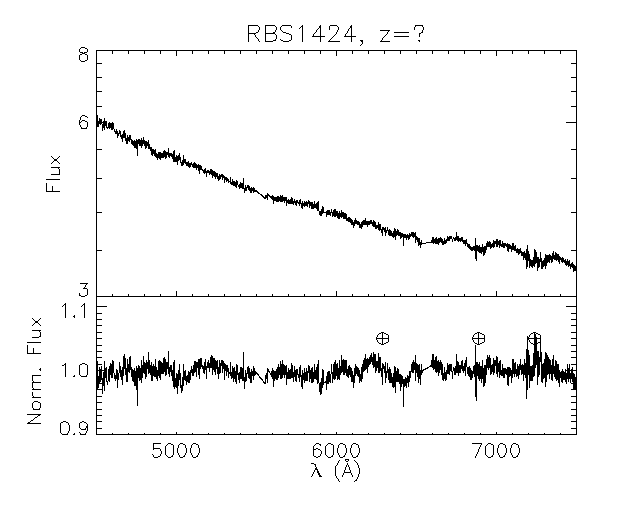}
 \includegraphics[width=6.8truecm,height=5.75truecm]{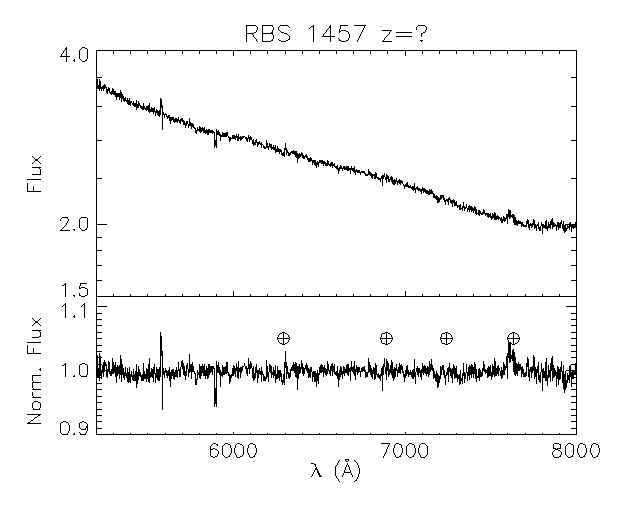}  \includegraphics[width=6.8truecm,height=5.75truecm]{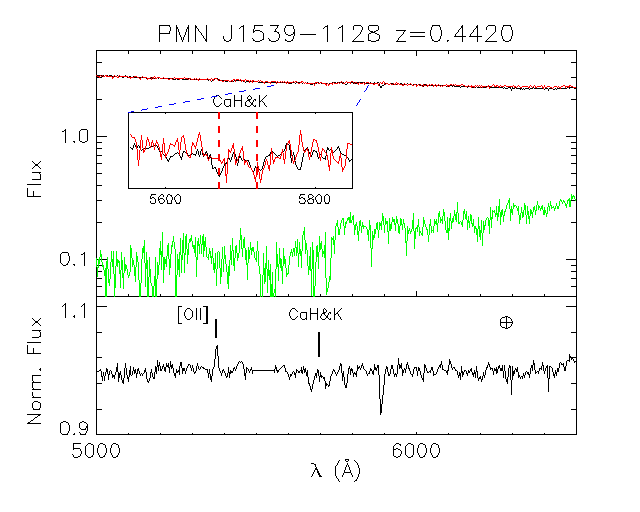}
  \includegraphics[width=6.8truecm,height=5.75truecm]{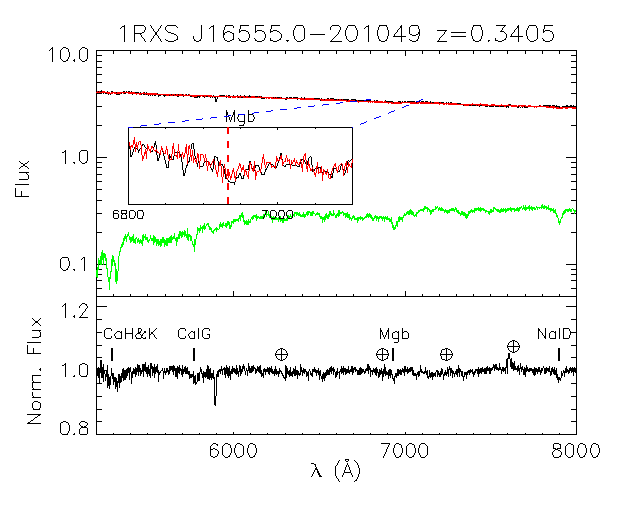}  
   \caption{Same as Fig. \ref{fig_spec1} for sources 9 to 16 in Table \ref{tabobs1}.}
\label{fig_spec2}
    \end{figure*}

A moderate S/N (72) spectrum obtained with CTIO/COSMOS for 3400 s has been taken by \citet{Des19}. We performed two observations of the source with VLT/FORS for a total exposure of 4790 s. An high S/N = 119 spectrum (Fig.~\ref{fig_spec2}, second row panel, on the left) allows us to detect several absorption features (Ca II HK, CaIG, Mg$_{b}$, Ca Fe, NaID), leading to a measured $z$ = 0.3171 $\pm$ 0.0002, compatible with the value presented in Marais \& van Soelen (in prep.). The host galaxy is quite luminous at $M_{R}$ = $-$23.2. 


\subsection{PKS 1424\texorpdfstring{$+$}{+}240 }

PKS\,1424+240 is a very bright $\gamma$-ray and TeV blazar, suggested as promising candidate neutrino emitter \citep{Aartsen20}. A featureless spectrum was reported by \citet{Shaw13}, while \citet{Fur13} determined a lower limit $z$ $\ge$ 0.6035 from the detection of hydrogen absorption systems in HST/COS spectra. A galaxy group at $z \sim$ 0.60 was identified in its environment and possibly associated with it \citep{Rov16}. Finally, weak detections of [OII] (EW $\sim$0.05 \AA) and [OIII]b (EW $\sim$0.1 \AA) emission lines at a S/N level between 3 and 5 \citep{Pai17b} allowed a redshift determination of $z$ = 0.604. While this redshift looks reliable, it would be useful to confirm it at a higher S/N level. Given that the above mentioned lines are of nebular origin and not variable \citep[see e.g.][]{Net90}, they should appear stronger when the non-thermal blazar flux declines. Thus an observation at a lower flux level of the AGN should confirm this result with higher S/N. We recently demonstrated the validity of this approach in Paper I by measuring the redshift of MAGIC J2001+435 through the detection of absorption features of the host galaxy during an optical low state.

\begin{figure}[htbp]
\centering
\includegraphics[width=6.8truecm,height=5.75truecm]{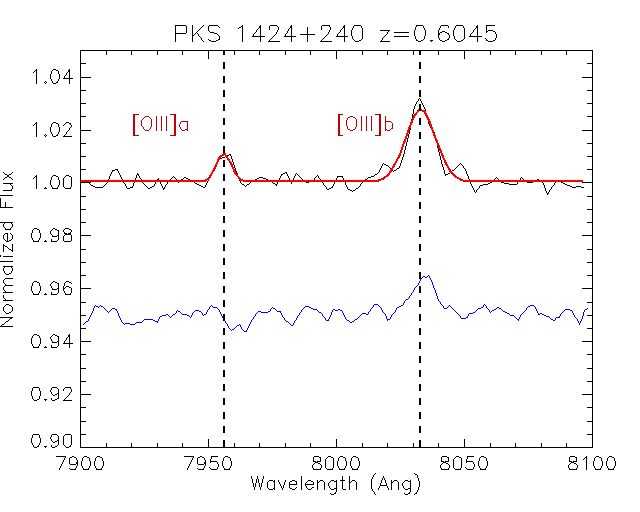}
\caption{[OIII] doublet during our observation of PKS\,1424+240 above and during \citet{Pai17b} observations. Note: the slight depression near the [OIII]a line in the \citet{Pai17b} spectrum corresponds to an uncorrected telluric feature.}
\label{fig_PKS_old-new}
\end{figure}

The observations from \citet{Pai17b} were performed on 2015 June 30 with the source at a magnitude R = 13.8, as obtained by the Tuorla monitoring. We detected an historical minimum of PKS\,1424+240 in March 2021 through the Tuorla monitoring: the flux level dropped to R = 14.70 $\pm $ 0.03, 0.9 magnitudes lower than the values obtained during previous observations. We observed PKS\,1424+240 on 2021 April 22 using OSIRIS on GTC (programme GTC02-21ADDT). Details are given in Table 1. The spectrum has a high S/N (305) and a power-law shape (Fig.~\ref{fig_spec2}, second row panel, on the right). In it, we detected [OII], [OIII]a and [OIII]b at EW = 0.15, 0.10,~and 0.35\,\AA~, respectively, with S/N ranging from 5 to 17 (see Table \ref{tabeqw2}). The lines are thus about three times more intense than the previous detection (Fig. \ref{fig_PKS_old-new}), which demonstrates the usefulness of low state spectroscopy for BL Lacs. The computed redshift is $z$ = 0.6045 $\pm$ 0.0002, slightly higher than the value in \citet{Pai17b}. The FWHM of the lines are 500 $\pm$ 33 km/s for [OII] and 400 $\pm$ 30 km/s for [OIII]a and [OIII]b. The redshift of PKS\,1424+240 is thus confirmed with higher S/N.

\subsection{RBS 1424}

Two featureless high S/N spectra obtained with VLT and GTC have been reported by \citet{Sbar06b} and \citet{Pai20}, respectively. We observed the source with SALT/RSS for 2390 s. Our high S/N ($\sim$ 130) spectrum (Fig.~\ref{fig_spec2}, third row panel, on the left) is featureless too. The redshift of the source remains undetermined.

\subsection{RBS 1457}

The  spectrum obtained  with the Hale Telescope at Mt. Palomar and presented in \citet{Shaw13} claims the detection of an MgII absorber at $\sim$ 3380 $\AA$, corresponding to a redshift lower limit of $z$ $\ge$ 0.21. \citet{Pir07} and \citet{Pai20} find featureless spectra, obtained with the ESO 3.6 m and the GTC, but their range does not cover the possible absorber.  We observed the source with VLT/FORS for 1950 s. The high S/N ($\sim$ 120) spectrum obtained (Fig.~\ref{fig_spec2}, third row panel, on the right), which starts at 5100 \AA~, is featureless too. As in the previous observations, due to the lack of coverage for the wavelength at which the MgII absorber should be detected, with our spectrum we cannot confirm or revise the redshift lower limit  suggested by \citet{Shaw13}.
 
\subsection{PMN J1539\texorpdfstring{$-$}{-}1128 }

Two intermediate S/N spectra are presented in \citet{Pen17,Gol21}, both of them featureless. We observed this BL Lac two times at separation of one day with SALT/RSS for a total exposure of 4380 s.  In our high S/N ($\sim$ 150) SALT/RSS spectrum (Fig.~\ref{fig_spec2}, bottom panel, on the left), we are able to detect Ca II HK and [OII]. We thus were able to determine the redshift of the source as  $z$ = 0.4420 $\pm$ 0.0002. The host galaxy is luminous with $M_{R}$ = $-$23.5.

\subsection{1RXS J165655.0\texorpdfstring{$-$}{-}201049  }

\citet{Jon09} reported a low S/N 6dF spectrum for this source, which has turned out to be featureless. A moderate S/N featureless spectrum obtained by SOAR/Goodman was reported in \citet{Pen17}. We observed it with VLT/FORS, obtaining a high S/N ($\sim$ 130) spectrum (Fig.~\ref{fig_spec2}, bottom panel, on the right). With our spectrum we were able to detect the Ca II HK, Mg$_{b}$, CaFe, and NaID absorption features and determine a redshift $z$ = 0.3405 $\pm$ 0.0002.

\subsection{TXS 1742\texorpdfstring{$-$}{-}078}

TXS\,1742$-$078 is a very absorbed source ($E_{B-V}$ = 0.802) near the Galactic plane (b = +10.8). A low S/N featureless spectrum obtained with the Hobby-Eberly Telescope has been presented in \citet{Shaw13}. The source has been observed with {\em Swift}-UVOT in the six UVOT filters and with SARA in SDSS g',r',i', and z' filters on 2018 August 28 (i'=17.97) but no photometric redshift or upper limit has been obtained by the SED fitting of the ten filters obtained by SARA+UVOT \citep{Raj20}. We noticed that at the time of the SARA observations the source was brighter (i'=17.97) with respect to the value reported by the Pan-STARSS survey (i=19). We observed the source twice within a week with VLT/FORS for a total exposure time of 5200 s. In the resulting moderate S/N =50 VLT/FORS spectrum (Fig.~\ref{fig_spec3}, top panel, on the left), we detected (with high S/N) the [OIII]b emission feature and with moderate S/N the [OIII]a and [OII] emission features, obtaining a redshift of $z$ = 0.4210 $\pm$ 0.0002.

\subsection{1RXS J184230.6\texorpdfstring{$-$}{-}584202}
  
A moderate S/N spectrum (S/N = 33) obtained using the 4 m telescope at Cerro Tololo InterAmerican Observatory in Chile with 3600 s resulted in a featureless spectrum \citep{Des19}. Observations with SOAR for 2200 s, with a S/N of 27 comparable to the spectrum obtain by \citet{Des19}, resulted to the detection of the Ca II HK absorption corresponding to a redshift $z$ = 0.421 \citep{Marchesini19}. Based on a moderate S/N spectrum obtained with NTT/EFOSC2 for 4500 s, in which the Ca II HK absorption has been detected, a tentative redshift of $z$ = 0.421 has been reported in Paper I. We took two VLT/FORS observations for a total exposure time of 4100 s in order to confirm or disprove the result. In the resulting high S/N = 77 spectrum (Fig.~\ref{fig_spec3}, top panel, on the right), we determined a redshift $z$ = 0.4226 $\pm$ 0.0003, based on the detection of the CaHK, CaIG, Mg$_{b}$, and CaFe absorption features.

\subsection{1RXS J203650.9\texorpdfstring{$-$}{-}332817 }

The redshift of the source is uncertain. A value of $z$ = 0.237 has been suggested by \citet{Alv16b} on the basis of a SOAR spectrum and the detection of the Ca II HK absorption. A low S/N NTT/EFOSC2 spectrum reported in Paper I did not offer the capability to confirm this detection. We observed it two times at a separation of two weeks with VLT/FORS for a total exposure time of 7200~s. The resulting high S/N spectrum (S/N = 143; Fig.~\ref{fig_spec3}, second row panel, on the left) is featureless, not confirming the redshift measurement of $z$ = 0.237.

\subsection{RBS 1751 }

 A possible MgII absorber at $z$ $\sim$0.681 was detected in a moderate S/N NTT/EFOSC2 spectrum presented in \citet{Gol21}. To confirm this redshift, we performed a short (350 s, S/N $\sim$ 20) VLT/FORS observation using Grism 1200B+97, confirming the absorber but at $z$ = 0.6185 $\pm$ 0.0001 (Fig.~\ref{fig_spec3}, second row panel, on the right).  We conclude that the blazar is at $z>$ 0.6185. The total EW of the system is 1.70 $\pm$ 0.14~\AA, and the two components have EW = 1.1 $\pm$ 0.1~\AA~and 0.6 $\pm$ 0.1~\AA. The ratio of the two components is 1.8 $\pm$ 0.3, which indicates a saturated system \citep[see e.g.][]{Spitzer78}.

\subsection{RX J2156.0\texorpdfstring{$+$}{+}1818 }

Two spectra of the source obtained with the Observatorio Astronómico Nacional San Pedro Mártir (OAN) for 5400 s \citep{Alv16} and SDSS \citep{Ahu20} are featureless. We took an high S/N ($\sim$ 90) spectra with Keck/ESI for 6000 s (Fig.~\ref{fig_spec3}, third row panel, on the left) and  we clearly detected a weak MgII absorber at $z$ = 0.6347 $\pm$ 0.0001 with a total EW of 0.5 $\pm$ 0.04~\AA. The two components' EWs are 0.3 $\pm$ 0.03~\AA~and 0.2 $\pm$ 0.03~\AA~, respectively, and their ratio is 1.5 $\pm$ 0.3, indicating mild saturation. Deep imaging observations detected the host galaxy of this object (Nilsson et al., in prep.), which suggests a very bright host.

\subsection{PMN J2221\texorpdfstring{$-$}{-}5224    }

\begin{figure*}[htbp!]
   \centering
\includegraphics[width=6.8truecm,height=5.75truecm]{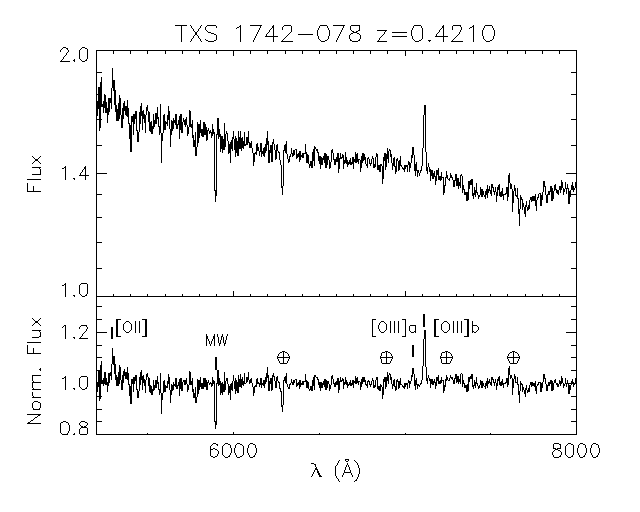}
 \includegraphics[width=6.8truecm,height=5.75truecm]{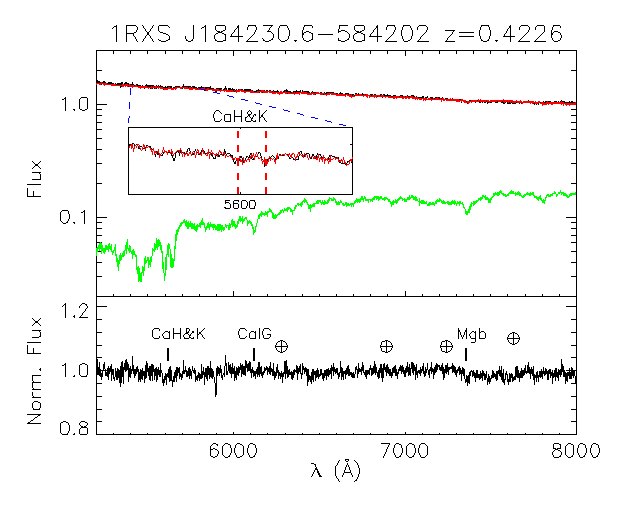}  \includegraphics[width=6.8truecm,height=5.75truecm]{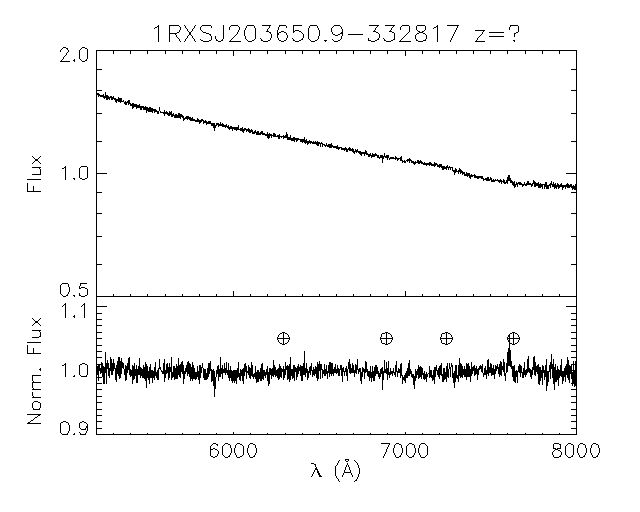}  
  \includegraphics[width=6.8truecm,height=5.75truecm]{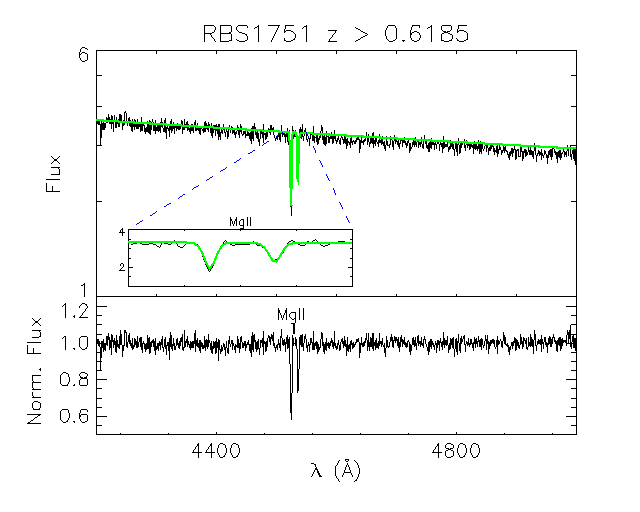}  \includegraphics[width=6.8truecm,height=5.75truecm]{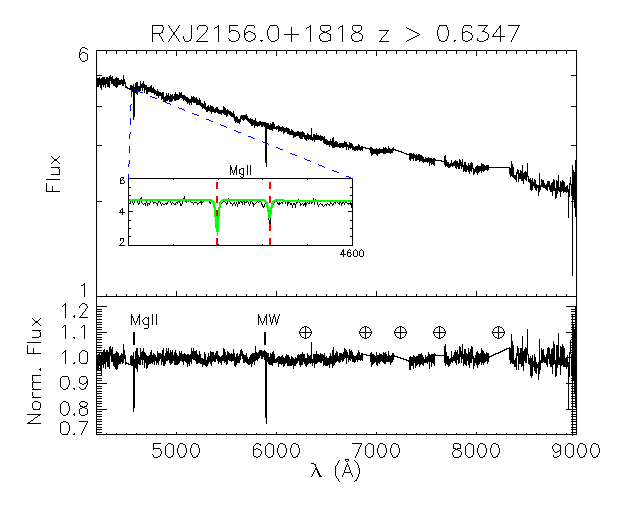}
 \includegraphics[width=6.8truecm,height=5.75truecm]{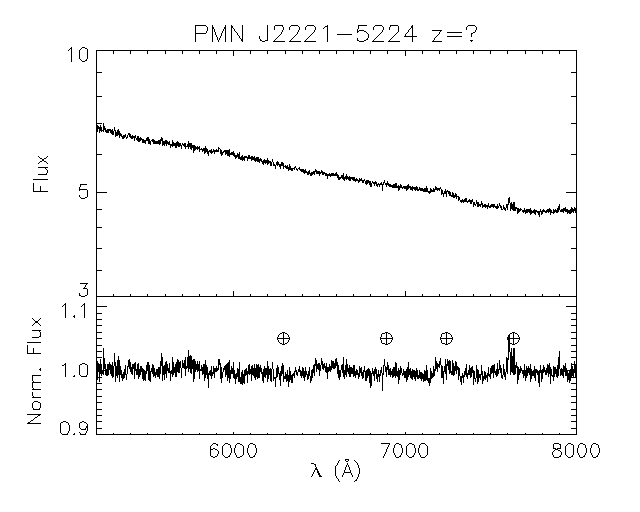}  \includegraphics[width=6.8truecm,height=5.75truecm]{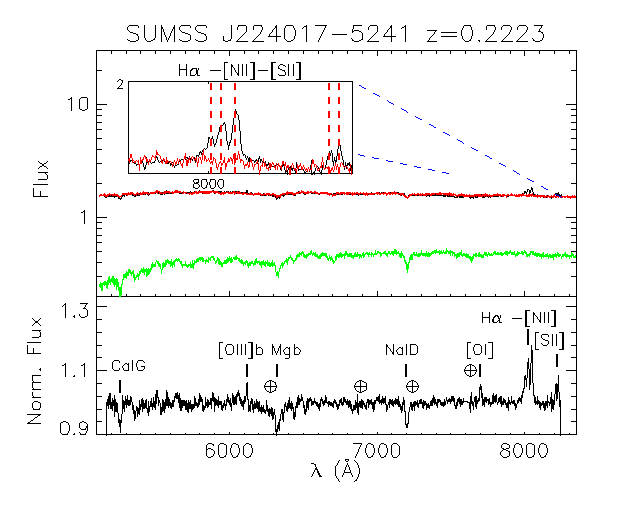}
  \includegraphics[width=6.8truecm,height=5.75truecm]{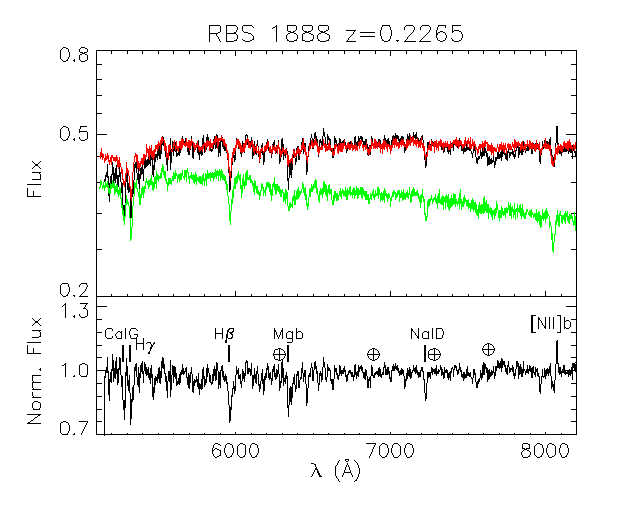}  
   \caption{Same as Fig. \ref{fig_spec1} for sources 17 to 24 in Table \ref{tabobs1}.}
\label{fig_spec3}
    \end{figure*}

A spectrum of PMN\,J2221$-$5224 was reported by \citet{Shaw13}. Using NTT/EFOSC2 they obtained a featureless spectrum with low S/N. We observed the source with VLT/FORS for 900 s, obtaining an high S/N ($\sim$ 150) spectrum (Fig.~\ref{fig_spec3}, third row panel, on the right) that is featureless. The redshift of the source remains undetermined.

\subsection{SUMMS J224017\texorpdfstring{$-$}{-}524111}

A low S/N spectrum of the source obtained by CTIO with 1950~s is reported by \citet{Des19}, resulting in a featureless spectrum. We observed it with VLT/FORS spectrum for  2600~s, obtaining a high S/N ($\sim$90) spectrum (Fig.~\ref{fig_spec3}, bottom panel, on the left). We detected several absorption (CaIG, Mgb, CaFe) and emission lines (see Table \ref{SUMSStab}) of the host galaxy at a common redshift of $z$ = 0.2223 $\pm$ 0.0001.

\subsection{RBS 1888}

For this source, \citet{Fis98} propose a redshift $z$ = 0.226 on the basis of a low S/N spectrum. To investigate this result we took an intermediate S/N ($\sim$60) spectrum with VLT/FORS with an integration time of 3900 s (Fig.~\ref{fig_spec3}, bottom panel, on the right). We detected several absorption features (CaIG, Mgb, CaFe, and NaID) and the fit of these features gives a redshift $z$ = 0.2265 $\pm$ 0.0002, thus validating the result presented in \citet{Fis98} with higher precision. The detection of strong H$\gamma$ absorption (EW = 2.8 $\pm$ 0.1 \AA) hints at a stellar population younger than what usually is observed in the blazar hosts \citep[see e.g.][ Paper I and II and references therein]{Falo14}.

\section{Comparison with ZTF and ASAS-SN light curves}

Spectroscopic observations triggered during a photometric minimum can improve the redshift measurement efficiency (see Section 6.12 of this paper and Paper I). This technique takes advantage of the weakening of the non-thermal emission during
the photometric minimum of the AGN, which allows for the emergence of the host galaxy features that are used to
determine the redshifts. For testing the importance of observing objects during a low-activity state, we investigated the light curves of the sources in Table \ref{tabobs1}. The light curves were 
collected from the Zwicky Transient Facility (ZTF) \citep{Masc19}\footnote{\url{http://www.ztf.caltech.edu}} and the All Sky Automated Survey for SuperNovae (ASAS-SN) \citep{Shap14,Koch17}\footnote{\url{https://asas-sn.osu.edu//}} in the period between January 2020 -- March 2022 to check if our spectroscopic measurements are performed during an high-, intermediate-, or low-activity state and the effect on the redshift measurement efficiency.  We briefly review the two survey characteristics. ZTF, which uses the 48-inch Oschin Schmidt telescope at Palomar Observatory, is sensitive down to r$\sim$20.6 (5$\sigma$ in 30 sec) and has a 1\arcsec\,pixel scale, but its coverage is limited mostly to the northern hemisphere. ASAS-SN is sensitive to g$\sim$ 18 (5$\sigma$ in 5 minutes) with a 8\arcsec\,pixel scale and, thanks to several sites around the globe, it covers the entire sky. Whenever available, we report ZTF measurements that have smaller uncertainties. 

We checked the light curves for consistency with archival photometric measurements and with our spectrophotometric magnitudes. This check led to the rejection of the ASAS-SN light curves of 1RXS\,J184230.6$-$584202 and SUMSS\,J113032$-$780105, which are strongly contaminated by nearby stars. Furthermore, the good-quality light curves of B2\,0557+38, 1RXS\,J085802.6$-$313043, and TXS\,1742$-$078 are not available due to their weakness. In Tables \ref{tabLC} and \ref{tabLC2}, the time separation between the spectroscopic measurement and the nearest observing data collected by ZTF and ASAS-SN, respectively, are reported together with the observed magnitude, the median, maximum, and minimum magnitude calculated over the entire period considered, and  the redshift or lower limit measured in this paper.
 
Taking a look at the ZTF light curves (Table \ref{tabLC}), at the time of our spectroscopic observations, 7 out of 11 sources have been observed at a magnitude comparable or higher (thus, a lower brightness) with respect to the median magnitude (hereafter, `low state') during 2020--2022. In particular, in the case of 1RXS\,J165655.0$-$201049 and RBS\,1888, the sources have been observed almost at the minimum magnitude. For five of these seven sources, a redshift has been determined with our new spectroscopic observations, the only exceptions being RBS\,1457 and RBS\,1751 (for which a high lower limit has been estimated). The other four sources have been observed at a lower magnitude (thus higher brightness) with respect to the median magnitude (`high state') and no redshift has been determined.

 In case of the ASAS-SN light curves, five out of eight sources have been observed in a low or intermediate (i.e. at a magnitude comparable to the median value) state, resulting in 2 redshift determinations. On the other hand,  three sources have been observed in an high state with respect of the median value resulting in one redshift determination. The only source for which the redshift has been measured in high state is SUMSS\,J224017$-$524111, which has a low jet-to-galaxy ratio (2.6 $\pm$ 0.2) and several emission lines, thereby simplifying the detection.

 In total, 12 sources in a low state (or intermediate in case of PMN\,J1539$-$1128 and SUMSS\,J082627$-$640414) yield 7 redshifts, while 7 sources in high state yield 1 redshift. As expected, `low state' observations are preferable for redshift determination.  We suggest that observers, when possible, should take into account this parameter when programming their spectroscopic observations.

\begin{table*}
\caption{\label{tabLC} Comparison with ZTF light curves (r band except 1RXS\,J013427.2+263846). Note: PMN\,J1539$-$1128 has two values corresponding to two different observations (see Table \ref{tabobs1}). Columns are: Source name; Delta time (days), which is the difference in time between our observation(s) date and the nearest ZTF observation; magnitude at nearest date; median, minimum, maximum magnitude of the source; and measured redshift. }
\centering
\begin{tabular}{lcccccc}
\hline\hline
Source name  & Delta time & Magnitude  & Median   & Maximum  & Minimum  & Redshift \\
              &  (days)   &                &            &            &    &   \\  
   (1)  & (2) & (3)    &  (4)  &  (5)   &  (6)  & (7) \\      
\hline 
1RXS\,J013427.2+263846  & 7.50  & 16.7$^*$  & 16.9  & 16.7  & 17.4 &  --- \\
1RXS\,J035000.4+064053 & 0.34  & 18.6  & 18.5  & 18.0  & 18.8 &  0.2730 \\
WISE\,J074627.03$-$022549.3  & 3.31  & 17.7  & 18.3  & 17.4  & 18.8  &  --- \\
PKS\,1424+240 & 1.03& 14.9 & 14.8 & 14.5 & 15.3 &  0.6045 \\
RBS\,1424  &  0.21 & 17.2  & 17.5  & 17.0   & 17.6  &  --- \\
RBS\,1457  & 2.36  & 17.6  & 17.5  & 16.7   & 18.4  &  --- \\
PMN\,J1539$-$1128  & 1.68/0.69  & 17.7/17.7  & 17.7  &  17.5 &  17.9 &  0.4420  \\
1RXS\,J165655.0$-$201049  & 0.28  & 17.8  & 17.4  & 17.0  & 17.9 &  0.3405 \\
RBS 1751 & 11.75 & 17.6 & 17.9 & 17.5 & 18.1 & $\ge$ 0.6185 \\
RX\,J2156.0+1818 & 0.25  & 18.0  & 17.6  & 17.0  & 18.3  & $\ge$ 0.6347  \\
RBS\,1888 &  0.29 & 19.3  & 19.1  & 18.6 & 19.4 &  0.2265 \\
\hline
\end{tabular}
\tablefoot{$^*$ i magnitude.}
\end{table*}

\begin{table*}
\caption{\label{tabLC2} Comparison with ASAS-SN light curves (g band). Columns are: Source name; Delta time (days) is the difference in time between our observation date and the nearest ASAS-SN observation; magnitude at nearest date; median, minimum and maximum magnitude of the source; and measured redshift. $^{*}$ indicates the average time between the two observations.}
\centering
\begin{tabular}{lcccccc}
\hline\hline
Source name  & Delta time & Magnitude & Median   & Maximum  & Minimum & Redshift \\
              &  (days)   &                &            &            &   &    \\  
   (1)  & (2) & (3)    &  (4)  &  (5)   &  (6)  & (7) \\      
\hline 
1RXS\,J005117.7$-$624154    &  0.96  &  16.6 & 16.6 &  15.8 & 17.4 &  ---  \\
SUMSS\,J014347$-$58455      &  1.92* &  17.3 & 17.0 &  15.5 & 17.4 & 0.3902 \\ 
1ES\,0505$-$546             &  1.66  &  17.2 & 17.2 &  15.6 & 17.6 & ---  \\
WISE\,J053629.06$-$334302.5 &  0.69  &  15.9 & 16.2 &  15.2 & 16.7 & --- \\ 
SUMSS\,J082627$-$640414     &  0.16  &  15.9 & 15.9 &  14.0 & 16.8 & 0.3397 \\
1RXS\,J203650.9$-$332817    & 8.19/24.18 & 18.2/18.2 & 18.0 & 16.7 & 18.6 &  ---  \\
SUMSS\,J224017$-$524111     & 3.77 & 17.5 & 17.6 & 16.3 & 18.2 &  0.2223 \\
PMN\,J2221$-$5224           &  2.54  &  16.4 & 16.7 &  14.8 & 17.3 &  --- \\
\hline
\end{tabular}
\end{table*}

\section{Conclusions}

In this work, 41 BL Lacs, which have been included in the 3FHL catalogue and selected as potential candidates for future observations with CTAO, were observed with Keck II, Lick, SALT, VLT, and GTC telescopes. The 41 targets are divided in 24 taken with 8-m class telescopes and whose spectra have been described in detail and 17, which only have low S/N, featureless spectra taken with the Lick telescope and are not discussed in detail here.
Of the 24 sources, 12 spectroscopic redshift (values between 0.2223 and 0.7018) and 1 tentative redshift (0.6622) have been determined. We also measured two redshift lower limits, $z$ $>$ 0.6185 and $z$ $>$ 0.6347. Taking a look at the quality of the spectra collected, for 15 out 24 BL Lacs, the S/N of the spectrum has been higher than 100; but it is only for 6 of them that a redshift measurement has been obtained. For the remaining 9 objects, the S/N values range between 23 and 95, from which we measured 6 firm redshifts, 1 tentative redshift, and 2 lower limits. As already noted in Paper II, an high S/N of the spectrum is an important but not sufficient condition for obtaining a redshift measurement. In fact, the contribution of the non-thermal jet emission can play an even more important role for the detection. We checked the activity level of the AGN in our sample investigating the ZTF and ASAS-SN light curves at the time of our spectroscopic observations and we found a high level of efficiency in the redshift detection in case of a low- or intermediate-activity state of the AGN; and, thus, a lower contribution of the non-thermal jet emission to the continuum spectrum. The case of PKS 1424+240, in which the emission lines observed by GTC during a low activity state are three times more intense than the lines previously observed, allowing us to improve the redshift determination of the source and confirming the   importance of the optical state of the AGN.
Following this indication, we are monitoring the optical state of the sources in our sample with the Telescopi Joan Oro \citep[TJO;][]{TJO} and the Rapid Eye Mount \citep[REM;][]{Zerbi01, Covino04} in order to perform follow-up spectroscopic observations of the targets in a low state.

The 11 firmly detected host galaxies in this paper have an average magnitude of $M_R$ = $-$22.9, brighter than the values obtained in Paper I and II. All the 11 sources can be adequately fitted with a local elliptical template \citep{Man01}. Emission lines are detected for five objects, all of them with an equivalent width smaller that the 5\,\AA, limit historically used to separate FSRQ and BL Lacs. 

The redshift measurement efficiency for the 24 sources taken with 8-m class telescopes is about 54 per cent, comparable to the results in Paper I and Paper II. The median redshift obtained with the new observations $z_{\rm\,med}^{\text{III}}$ = 0.39 is comparable to the value reported in Paper II ($z_{\rm\,med}^{\text{II}}$ = 0.37) and higher than the value reported in Paper I ($z_{\rm\,med}^{\text{I}}$ = 0.21). We also computed the combined results of our programme to date (see Table \ref{resultsPapers}). To do this, we took into account repeated observations. In this paper, we re-observed PKS\,1424+240, PMN\,J1539\texorpdfstring{$-$}{-}1128, 1RXS\,J184230.6\texorpdfstring{$-$}{-}584202 (confirming its tentative redshift reported in Paper I), 1RXS J203650.9\texorpdfstring{$-$}{-}332817, and RBS\,1751 (confirming its tentative lower limit reported in Paper I). Thus, we have two tentative redshifts remaining (B2\,0557+38 from this paper and RX\,J0819.2+0756 from Paper II) and no tentative lower limits remaining.
Up to now, we observed 63 independent targets and obtained 37 solid redshifts with an efficiency of about 59\% and a median redshift $z_{med}^{\text{tot}}$ = 0.30.

By considering the high efficiency of redshift determinations in cases of low activity of the AGN, re-observations of sources for which a featureless spectrum has been obtained, despite a S/N $>$ 100 spectrum, will be performed during a low activity state. For a better characterization of our sample of BL Lac objects, we are also currently performing deep near-IR imaging of these sources (Nilsson et al., in prep.). The most distant host galaxy we detected is that of RX\,J2156.0+1818, for which we report a spectroscopic lower limit of 0.6347 here. 

 With the new spectroscopic observations performed here, we obtained 9 redshifts and 1 tentative redshift higher than 0.3, similar to the numbers obtained in paper II (i.e. 8 redshift and 1 tentative redshift with $z$ $>$ 0.3), but significantly higher than the numbers obtained in paper I (i.e. 2 redshifts and 1 tentative redshift with $z$ $>$ 0.3).  In total, 19 redshifts and 1 tentative redshift with $z$ $>$ 0.3 have been determined in our programme. As a comparison, 9 BL Lacs with redshifts of $z$ $>$ 0.3 are reported in TeVCat presently. Our sources thus represent an important pool of possible TeV sources at $z$ $>$ 0.3 that can be targets of VHE ground-based observatories such as CTAO. These objects are important for determining the luminosity function of BL Lacs and for EBL studies. For this reason, this redshift measurement campaign is recognized as necessary to support for future CTAO observations of AGN, in particular, for CTAO Key Science Projects on AGNs \citep[e.g.][]{CTAcons20,CTACons21-prop}. 

\begin{table*}
\caption{\label{resultsPapers} Number of observed sources, along with redshift and lower limit measurements (uncertain results given in parentheses) for different groups of sources and for the whole sample for Paper~I, Paper~II, this paper (Paper~III), and the combined results of our programme. Note: the combined results take into account repeated observations (see discussion for details). The efficiency numbers reported here for Paper~III are the ones of the sample observed by 8-m class telescopes. C: These uncertain results are confirmed in Paper III.}

\centering
\begin{tabular}{llllllllll}
\hline\hline

Paper & Number of & Redshifts & Redshift & S/N $\geq$ 100 & S/N < 100 & $z_{med}$ & Efficiency \\
& targets & ($z$) & lower limits & sources ($z$) & ($z$) &  & \\
(1)  & (2) & (3)    &  (4)  &  (5)   &  (6)      &  (7)   &  (8)  \\  

\hline

I  & 19/19 & 11 (+1)$^{C}$ & 2(+1)$^{C}$ & 9~~(8) & 10 (3+1) & 0.21  & 11/19 ~~~~~~~| 58\%\\
II & 25/33 & 14 (+1) & 2 & 7~~(1)& 26 (13+1) & 0.37  & 14/25 (33) | 56\%  \\
III & 24/41 & 12 (+1) & 2 & 15 (6) &  26 (6+1) & 0.39  &  12/24 (41) | 54\% \\
\hline
\hline
Combined & 63/83 & 37(+2) & 6 & 31(15) & 62(22+2) & 0.30 & 37/63  | 59\% \\
\hline
\end{tabular}
\end{table*}

\begin{acknowledgements}
This paper went through internal review by the CTA consortium. We thank M. Cerruti and N. Masetti for their helpful comments and suggestions as internal CTA reviewers. This research has made use of the CTA instrument response functions provided by the CTA Consortium and Observatory, see \url{https://www.cta-observatory.org/science/cta-performance/}(version prod3b-v1) for more details. We gratefully acknowledge financial support from the agencies and organisations listed here: \url{http://www.cta-observatory.org/consortium_acknowledgments}, and in particular the U.S. National Science Foundation Grant PHY-2011420.
 The authors wish to recognise and acknowledge the very significant cultural role and reverence that the summit of Maunakea has always had within the indigenous Hawaiian community.  We are most fortunate to have the opportunity to conduct observations from this mountain.
Some of the observations reported in this paper were obtained with the Southern African Large Telescope (SALT). Based on observations made with the GTC telescope, in the Spanish Observatorio del Roque de los Muchachos of the Instituto de Astrofísica de Canarias, under Director's Discretionary Time. This research has made use of the SIMBAD database, operated at CDS, Strasbourg, France. We thanks the anonymous referee for careful checking the manuscript and providing useful comments and suggestions.

\end{acknowledgements}

\bibliographystyle{aa} 
\bibliography{ctaredshift}        

\begin{appendix}

\section{`Additional sample' observations with the Lick telescope}\label{app}

 Additional observations performed with Lick/KAST are reported here. They yield S/N < 100 and no detection of spectral features for a total of 26 spectra for 18 blazars. One of them (1RXS\,J035000.4$+$064053) has an high S/N spectrum obtained with Keck/ESI, reported in Table\ref{tabobs1}.

\begin{sidewaystable*}
\small
\caption{\label{tabobs2} Analysis results on 26 featureless spectra of 18 blazars observed with Lick/KAST. They include one spectrum of 1RXS J035000.4+064053 for which we report Keck/ESI observations in Table \ref{tabeqw} which resulted in its redshift measurement: $z$ = 0.2730. The Lick/KAST spectrum has a S/N that is too low to confirm this result.}
\centering
\begin{tabular}{lcccclllllll}
\hline\hline
3FHL name                &   4FGL Name             & Source name                                          & RA             & Dec                & Start Time                  & Exp.  & Airm.& Seeing     & S/N & Slope & R$_{\rm c}$ \\
                                   &                                   &                                                                &                    &                      &  UTC                            &  (sec) &       & (\arcsec)   &         &           &        \\  
 (1)                             & (2)                              &             (3)                                               &  (4)             & (5)                  &     (6)                           &     (7)  & (8) &   (9)           &(10)    & (11) &    (12)        \\ 
\hline
3FHL J0131.1+6120 & 4FGL J0131.1+6120      & 1RXS J013106.4+612035$^{\dagger}$  & 01 31 07.2 & +61 20 33   &  2021-09-02 11:03:58 & 2700 & 1.10 & 1.7 &   7    &    -1.3$\pm$0.2     &    17.5$\pm$0.2        \\
                                  &                                        &                                                                &                   &                    & 2021-11-11 04:39:34 & 5400 & 1.11 & 1.8 &    9     & -1.5$\pm$0.2        &   17.9$\pm$0.2        \\
                                  &                                        &                                                                &                   &                    & 2021-12-02 02:22:55 & 3600 &1.17 & 1.4 &   13      & -1.8$\pm$0.8        &  17.6$\pm$0.3         \\
3FHL J0318.8+2135 & 4FGL J0318.7+2135     & MG3 J031849+2135                               & 03 18 45.7 & +21 34 37    & 2021-11-11 06:17:47  & 7200 & 1.07 & 2.0 & 55         & -0.7$\pm$0.1        & 16.2$\pm$0.2           \\
                                  &                                        &                                                                &                   &                    & 2021-12-02 03:28:46  & 5400 & 1.26 & 2.0 & 31        & -0.5$\pm$0.1         & 16.6$\pm$-0.1           \\
3FHL J0322.0+2336 & 4FGL J0322.0+2335     & MG3 J032201+2336$^{\dagger}$           & 03 22 00.  &  +23 36 11    & 2021-11-11 08:24:16 & 7200 & 1.08 & 1.4 &    42     &    -1.0$\pm$-0.1      &  16.4$\pm$0.2         \\
3FHL J0350.0+0640  & 4FGL J0350.0+0640     & 1RXS J035000.4+064053$^{\dagger}$ & 03 49 57.8 & +06 41 26    & 2021-01-09 03:20:05 & 3335 & 1.19 & 3.2 &     2    & -0.8$\pm$2.0        &  19.2$\pm$0.2         \\
3FHL J0422.3+1949  & 4FGL J0422.3+1951   & SHBL J042218.4+195051$^{\dagger}$    & 04 22 18.0 & +19 50 54   & 2021-12-02 07:14:03 & 10800 &1.10 & 1.6 &    8              & -1.5$\pm$0.2         & 18.2$\pm$0.3         \\
3FHL J0423.8+4149   & 4FGL J0423.9+4150   & 4C +41.11$^{\dagger}$                            & 04 23 56.1 & +41 50 0.3 & 2021-11-11 10:31:24  & 3600 & 1.07 & 1.7 &  5       &  -1.4$\pm$0.7        &  17.9$\pm$0.2          \\
                                   &                                      &                                                                 &                   &                   & 2021-12-02 05:06:00  & 7200 & 1.08& 2.0 &   5     & -1.4$\pm$0.7       &   17.9$\pm$0.2        \\         
3FHL J0434.7+0921  & 4FGL J0434.7+0922      &   TXS 0431+092                                    & 04 34 40.9 & +09 23 48   & 2021-01-09 04:45:57  & 7200 & 1.15 & 2.4 &  16       & -1.0$\pm$0.1         &   16.9$\pm$0.2         \\
3FHL J0515.8+1528.  & 4FGL J0515.8+1527    & GB6 J0515+1527$^{\dagger}$             & 05 15 47.3 & +15 27 16. & 2021-12-02 10:23:59 & 7200 & 1.37 & 1.5 &    21     & -1.0$\pm$0.2    &  16.6$\pm$0.2          \\
3FHL J0540.5+5823  & 4FGL J0540.5+5823     &   GB6 J0540+5823$^{\dagger}$            & 05 40 30.0 & +58 23 38   & 2021-01-09 07:00:33 & 8900 & 1.11 & 2.6 & 34      &  -0.8$\pm$0.2         & 16.0$\pm$0.2           \\
3FHL J0706.5+3744.  & 4FGL J0706.5+3744   &   GB6 J0706+3744$^{\dagger}$              &  07 06 31.7   & +37 44 36 & 2021-11-11 11:37:28 & 5400  & 1.01 & 1.7 &  33       &  -0.8$\pm$0.1        &   16.9$\pm$0.2        \\
                                  &                                        &                                                                &                      &                  & 2021-12-02 12:31:14 & 3600 & 1.15 & 1.6 & 24        &  -0.7$\pm$0.2       &  17.4$\pm$0.2
                                                                 \\
3FHL J0905.5+1357  & 4FGL J0905.6+1358     &   MG1 J090534+1358                             & 09 05 35.0 & +13 58 06   & 2021-01-09 09:38:58 & 6800 & 1.11 & 3.1 & 46       &    -1.1$\pm$0.1      &  16.1$\pm$0.2         \\
                                  &                                        &                                                                &                    &                   & 2021-04-17 04:24:49 & 7200 & 1.19 & 2.5 &   46      &  -1.0$\pm$0.1       &  16.1$\pm$0.2         \\
3FHL J1037.6+5711   & 4FGL J1037.7+5711    &   GB6 J1037+5711$^{\dagger}$             & 10 37 44.3 & +57 11 56   & 2021-01-09 11:46:21 & 7000 & 1.09 & 3.4 &  54       &  -0.6$\pm$0.1       &  15.2$\pm$0.1         \\
3FHL J1150.5+4154   & 4FGL J1150.6+4154     &   RBS 1040$^{\dagger}$                       & 11 50 34.8 & +41 54 40   & 2021-01-09 13:50:54 & 1750 &1.03 & 2.8 & 34        & -1.2$\pm$0.1        &   15.9$\pm$0.2        \\
                                    &                                      &                                                              &                   &                & 2021-04-17 08:47:49  & 5340 & 1.21 & 2.7 & 46         &  -0.9$\pm$0.1       &   15.9$\pm$0.2         \\
3FHL J1233.7$-$0145 &  4FGL J1233.7$-$0144 &   NVSS J123341-014426$^{\dagger}$  & 12 33 41.3 & -01 44 24   & 2021-04-17 06:33:53 & 7200 & 1.31 & 2.9 &     8    &   -0.5$\pm$0.2        &     17.9$\pm$0.1         \\
3FHL J1811.3+0341   &  4FGL J1811.3+0340    &   NVSS J181118+034113$^{\dagger}$    & 18 11 18.1 & +03 41 14  & 2021-04-17 10:28:50 & 6360 & 1.33 & 2.6 &  53       & -0.9$\pm$0.1       &  15.4$\pm$0.2      \\
3FHL J1925.0+2815   &  4FGL J1925.0+2815   &   NVSS J192502+28154                          & 19 25 02.2 & +28 15 42  & 2021-06-09 09:54:09 & 4200 & 1.01& 3.6 &   13      &   -1.2$\pm$0.6      &  15.0$\pm$0.2         \\
                                  &                                        &                                                                &                     &                  & 2021-09-02 04:08:09 & 7200 & 1.02 &1.8 & 33        &   -0.2$\pm$0.1       &  16.9$\pm$0.1        \\
3FHL J2247.9+4413  &  4FGL J2247.8+4413    &   NVSS J224753+441317$^{\dagger}$    & 22 47 53.2 & +44 13 15 &  2021-09-02 06:17:07  & 5400 & 1.04 & 1.7 &  32       &   -1.4$\pm$0.3       &  16.6$\pm$0.3         \\
3FHL J2304.7+3705  &  4FGL J2304.6+3704   &   1RXS J230437.1+370506$^{\dagger}$  & 23 04 36.7 & +37 05 07 &   2021-09-02 09:51:54 & 2720 & 1.08 & 1.8&        22  &  -1.2$\pm$0.2       &  17.1$\pm$0.2         \\

\hline\hline
\end{tabular}
\tablefoot{The columns contain:  (1) 3FHL Name, (2) 4FGL Name, (3) Source Name, a $^{\dagger}$ indicates the source is in BZCat, (4) Right Ascension (J2000), (5) Declination (J2000), (6) Start Time of the observations, (7) Exposure Time, (8) Average Airmass, (9) Average Seeing, (10) Median Signal-to-Noise per bin measured in continuum regions, (11) Power-Law Slope with errors, (12) R$_{\rm c}$, Cousins magnitude of the BL Lac spectrum corrected for reddening, telluric absorption, and slit losses with errors. }
\end{sidewaystable*}

\end{appendix}

%
%


\end{document}